\documentstyle[multicol,prl,aps,epsf]{revtex}
\begin{document}
\draft
\title{
Spontaneous alloying in binary metal microclusters\\ 
- A molecular dynamics study -
}
\author{Yasushi Shimizu\thanks{e--mail:
shimizu@ifc.or.jp}}
\address{
Institute for Fundamental Chemistry, Takano-Nishihiraki-cho 34-4,
Sakyou-ku,Kyoto 606-8103, Japan
}
\author{Kensuke S.Ikeda, 
\thanks{e--mail: ahoo@mp0tw009.bkc.ritsumei.ac.jp}}
\address{
Department of Physical Sciences, 
Ritsumeikan University, Noji-higashi 1-1-1, Kusatsu 525-8577, Japan
}
\author{
Shin-ichi Sawada, 
\thanks{e--mail: s-sawada@kwansei.ac.jp}}
\address{
School of Science, Kwansei Gakuin University,
1-155, Uegahara, Nishinomiya 662-0891, Japan
}
\date{August 23, 2000}
\maketitle
\begin{abstract}
Microcanonical molecular dynamics study of spontaneous alloying(SA), 
which is a manifestation of fast atomic diffusion 
in a nano-sized metal cluster, is done in terms 
of a simple two dimensional binary Morse model.
Important features observed by Yasuda and Mori are well reproduced 
in our simulation.
The temperature dependence and size dependence of 
SA phenomena are extensively explored by 
examining long time dynamics. 
The dominant role of negative heat of solution in completing 
SA is also discussed. 
We point out that a presence of melting surface 
induces the diffusion of core atoms even if they are solid-like.
In other words, the {\it surface melting} at substantially low 
temperature plays a key role in attaining SA.
\end{abstract}
\begin{multicols}{2}
\section{Introduction}
Microclusters exhibit neither the properties of 
bulk nor those of molecules.
One of the most important features of atomic and molecular clusters is
the presence of the large portions of surface atoms, which provide 
large fluctuation in their motion. 
A lot of interesting static and dynamical properties 
of nano-sized clusters are found by many authors during the last 
two decades\cite{Sugano}. 
If we restrict ourselves to dynamical aspects 
of microclusters, for instance, 
it was shown that small metal clusters fluctuate 
between different multiply twinned and single-crystal structures rather 
than having fixed structures\cite{Iijima}.
From the viewpoint of equilibrium thermodynamics, Ajayan pointed out
that this is a manifestation of a quasi-molten state 
where the Gibbs free energy surface as a function of the cluster 
morphology is quite shallow\cite{AandM}. 
From the viewpoint of a molecular dynamics(MD) study, on the other hand, 
Sawada and Sugano showed that the structural 
change of Au clusters observed in experiments is regarded as 
a floppy motion between local minima of a potential surface due to 
the dynamical nature of clusters\cite{Sawada}.
In both cases we may say small clusters suffer from anomalously 
large dynamical fluctuations. 
Owing to the presence of such large fluctuations, 
as pointed out by Sugano, it is hard to 
give a clear answer for the following naive questions;
Are microclusters like solids where atoms are oscillating around 
their respective equilibrium positions? Are they like liquids where 
atoms move diffusively? Or, are they fluctuating between different solid 
phases during the course of their motion?\cite{Sugano}
In fact, according to the works by  Berry and his coworkers 
in their microcanonical MD study of a Lennard-Jones cluster, 
there exists the intermediate phase called 'co-existing phase' of liquid 
and solid\cite{Berry1}.
While numerical searches of stable structures of small
microclusters have been done extensively, research  
is left untouched on extremely long time dynamics 
beyond micro seconds, which is responsible for diffusion process
in clusters.  
In the present paper, we discuss a novel fast diffusion 
process which was experimentally discovered by Yasuda and Mori(YM) 
in nano-sized binary metal
clusters, because  it is a manifestation of an anomalous
diffusion process peculiar to a microcluster\cite{YandM}.
The aim of the present work is to realize spontaneous 
alloying(SA) with a concise 
model and to elucidate what kind of dynamics dominates SA 
with an extensive numerical study.\par 
The present paper is organized as follows.
In the next section we mention the experimental results of SA 
and an outline of our motivation.
In Sec.III our model for the MD simulation of SA is introduced with 
a physical assumption which we made to prepare appropriate 
initial configurations.
Numerical results and observations of the MD simulation 
are presented in Sec.IV.
Some useful quantities are introduced 
to characterize atomic fluctuation and rearrangement in a cluster. 
A comparison between experimental and numerical results are 
discussed in Sec.V. 
In Sec.VI the differences between surface and core atoms in a 
cluster are emphasized by paying attention to the activity of atoms 
manifested by their fluctuation and rearrangement. 
A special emphasis 
is put on the important role of surface melting in SA. 
Lastly, in Sec.VII we briefly make a concluding statement regarding 
the interpretation of the results. 
\section{Summary of YM's experimental results and an unusual 
feature of SA} 
\subsection{YM's experimental results}
In 1992 a novel transport phenomenon in a 
nano-sized metal alloy cluster was reported by YM\cite{YandM}.
By using an evaporator they deposited individual solute atoms(copper) on 
the surface of host nano-sized clusters on amorphous carbon film 
at room temperature and observed the alloying behavior with 
a transmission electron microscope. 
In Fig.1(a) their {\it in  situ} observation is schematically described. 
In \cite{YandM} they demonstrated that gold clusters 
promptly changed into highly 
concentrated, homogeneously mixed $(Au-Cu)$ alloy clusters. 
This process is termed as {\it spontaneous alloying}(SA).
SA is similarly observed in many nano-sized binary clusters, such as 
$(Au-Ni)$,$(In-Sb)$,$(Au-Zn)$, and $(Au-A\ell)$\cite{YandM}.
They examined the presence and absence of SA 
for clusters of various sizes.
\begin{figure}
\begin{center}
\epsfysize=4cm
\epsfbox{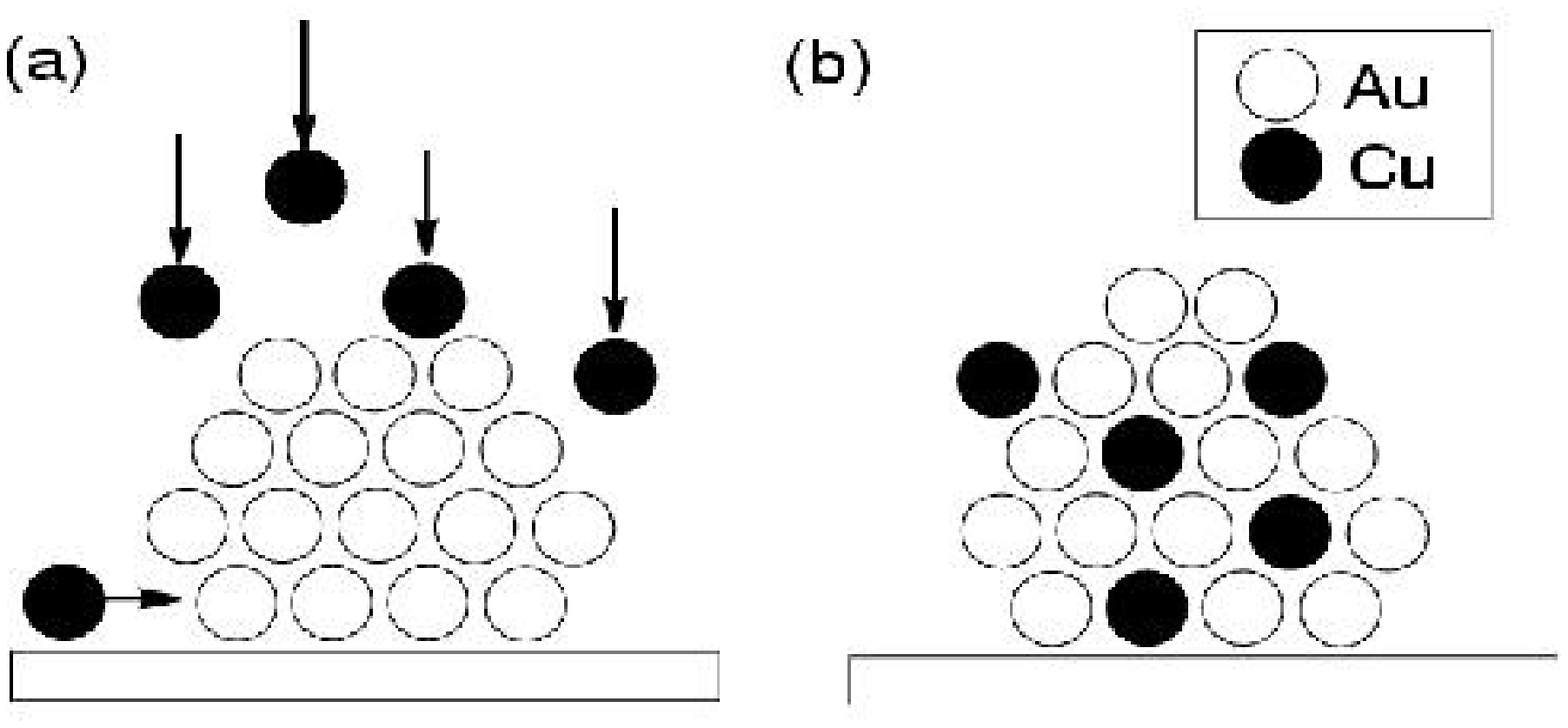}
\end{center}
{\bf Fig.1} A schematic picture of the 
{\it in  situ} observation of SA by YM. 
White and black circles 
denote gold and copper atoms, respectively.(a) Before the onset of SA, 
individual copper atoms are deposited on a gold cluster. (b) After SA 
is completed, copper atoms  dissolve into a gold cluster to form a 
homogeneously mixed alloy cluster. 
\end{figure}
They concluded that the remarkable features of SA phenomenon 
can be summarized as follows:\\
(1)The diffusion rate of copper atoms in clusters is approximately 
9 orders of magnitude faster than that in bulk crystalline alloys.
By use of the simple relation, $x=\sqrt{Dt}$, between the diffusion 
coefficient D and the time t, needed to achieve diffusion of solute 
atoms across the distance $x$, YM roughly evaluated the value 
$1.1\times 10^{-19}[m^2/s]$ as a lower limit. 
Note that the diffusion constant of copper 
in the bulk gold is known to be $2.4\times 10^{-28}[m^2/s]$ at $300[K]$ 
\cite{Kuba}.\\
(2) Negative heat of solution plays an important 
role in 
enhancing and supressing the SA process.
Indeed, SA has never been observed
in the combination of the solute and host atoms
with sufficiently large positive heat of solution. 
However, it is worth mentioning that SA occurs even with 
the species with almost null heat of solution. \\
(3)Temperature is also an important factor controlling SA.
At relatively high temperature($T\sim 245$[K]), Cu atoms 
can dissolve well into the center of a 4nm-sized Au cluster, 
whereas at medium temperatures($T\sim 215$[K],$165$[K]), 
the dissolution of copper takes place only over a limited, 
shell-shaped region beneath the surface of a 4-nm sized cluster, 
and the thickness of that region where a solid solution is 
formed decreases with the decrease in temperature.\\ 
(4) With increase in cluster size the occurrence of rapid SA is
 suppressed. In Au clusters of approximately 10nm in the mean size, 
rapid alloying of Cu takes place only at the shell-shaped region beneath 
the free surface of an individual cluster and pure gold was retained at 
the central region of the cluster.  In Au clusters of approximately 30nm in 
the mean size, no rapid alloying of Cu does not take place. It should be
stressed that the critical size of the SA increases with the negative
heat of solution and temperature. \\ 
(5) SA takes place in a solid phase, which was confirmed by the
fact that no changes were observed in  
the multiply twinned structure of the host cluster during the alloying process. 
Although the electron beam heating seemingly brings considerable influence on
SA, the estimated magnitude of the temperature rise in 
atom clusters was of the order of $10[K]$, which causes no
significant effect on SA.\\
\subsection{What is unusual in SA?}
The experimental results mentioned above 
suggest the presence  of an unexpectedly 
fast diffusion process, which is controlled by negative 
heat of solution and temperature, in a small sized cluster. \par
It is worth while to state {\it what is unusual} in SA. 
It is quite natural for Au and Cu atoms to mix and 
and change into an alloyed state, which is entropically 
preferable from the viewpoint of equilibrium statistical 
mechanics. 
In addition, negative heat of solution 
ensures that the alloyed state is enthalpicaly preferable. 
In fact near icosahedral-like ground states, mixed structures are 
preferred over the segregated ones for bimetallic clusters with  
$13-14$ atoms\cite{Marcos}.
Thus it seems to be no surprise to find out 
a {\it spontaneously} alloyed state in a binary cluster. 
The really unusual point of SA is 
that the alloying completes at least 
within the time scale of {\it second} even at room temperature.
How is this diffusion realized? 
What we would like to elucidate is the atomistic mechanism 
of such a fast diffusion.
Before pursuing this question, it is helpful to 
confirm how fast the diffusion is.
Supposing that a diffusion process of 
impurity atoms obeys Arrhenius law, YM's experimental result implies that 
the activation energy is effectively lowered, at least, 
by $40-50\%$ in Au clusters compared with bulk crystalline Au.  
It is important to note that the diffusion coefficient of atoms  
on a clean surface is many orders of magnitude faster than that in bulk. 
In fact the activation energy of the diffusion constant 
on a clean surface is about $35\%$ of that in the bulk\cite{Guiraldenq}.
That is, the diffusion coefficient of Cu atoms in 
a nano-sized cluster is almost 
comparable to that on a surface.
Atoms are easy to move on a cluster surface, 
because the surface is populated by point defects.
Considering that the fraction of surface atoms  
in a cluster becomes significant as the cluster size decreases,
it is clear that the rapid surface 
diffusion is relevant for the rapid SA phenomenon. 
On the other hand, it is important to note 
that SA is a manifestation 
of atomic movement in the {\it radial} direction of the cluster.  
How is the fast diffusion {\it parallel} to the surface of cluster related
to the rapid diffusion in the {\it radial} direction?
A remarkable difference between a bulk surface and a cluster surface 
is that the latter has a non-vanishing curvature, 
which makes the surface more deformable.
An easily deformable surface may influence the
manner of atomic diffusion along the surface and may
drastically accelerate the radial diffusion rate\cite{SM}.
%
How is such a rapid diffusion  enhanced or suppressed by the 
key factors such as  magnitude of negative heat of solution, cluster
size and temperature rise? \par
In our previous work, we reported some 
preliminary results which included the numerical simulation of SA by
using a 2-dimensional(2D) binary Morse model\cite{SIS}. 
We pointed out that
some features of the experimental results summarized by (1)-(5) 
may be reproduced.  However, the results presented there were not
complete. The time scale of simulation was not long enough to reproduce 
the whole SA process in larger-sized clusters and the tempareture
of simulation was not lower enough than the melting temperature. 
In particular, the dependence of the alloying process on the cluster 
size, which is the heart of YM's experiments, was not 
made clear. 
The aim of the present paper is to show systematic results
of a further extensive and comprehensive numerical simulation
and demonstrate how our simple-minded Morse model of clusters
reproduces the essential feature of YM's experiment. 
A particular emphasis is put on the fluctuation and rearrangement
of cluster atoms, which contain useful information for 
elucidating the atomistic mechanism of the SA process. 
\section{A numerical approach: How to prepare a model with appropriate 
initial conditions}
\subsection{A model cluster} 
There are several proposed empirical potentials 
which successfully mimic the equilibrium properties of bulk 
metals such as a lattice constant, 
bulk modulus, elastic constants and  sublimation energy.
Among them the so-called Embedded Atom Method(EAM) is 
a well-known model for alloy systems\cite{EAM}.  
However, it has 
many parameters which should be well-adjusted to yield plausible values 
for equilibrium properties of bulk metals.  
Moreover, transferability or applicability of these potentials 
to a cluster system is still unknown\cite{Rey}. 
Because a simpler model is better to get physical insight into 
the mechanism of SA, 
we employ a Morse model which has can explain 
qualitative aspects of the experimental results\cite{SIS,COM}.\par
More specifically, 
we choose the Morse model for the two reasons:\\
(1)Unlike EAM, the heat of solution, which is the key parameter
of our simulation, can easily be controlled by 
a single parameter. EAM has many parameters which 
is influential in changing heat of solution.\\
(2)By using pairwise potential we can considerably 
reduce the simulation time.\\
Furthermore, in the present work we use a 2D Morse cluster
rather than the realistic 3D cluster. 
The reasons why we examine the 2D model are twofold. 
First, the computation
time for the 2D model is much shorter than that for the 3D model.
In realistic 3D simulations corresponding to the experimental 
condition, time evolution of more than 1000 atoms should be traced 
for longer than a few microsecond. 
This is still beyond recent computational capability. 
Secondly, visualization of individual atomic motion in the 
SA process can more easily be done with  
2D model than that with 3D model.\par
We take 2D Morse potential Hamiltonian
\begin{eqnarray}
H=\sum_{i=1}^{N}\frac{1}{2m}({p_x^{(i)}}^2 +{p_y^{(i)}}^2)
+\sum_{i<j}V_{kl}(r_{ij}), 
\end{eqnarray}
and 
\begin{eqnarray}
V_{kl}(r)=
\epsilon_{kl} \{ e^{-2\beta_{kl}(r-r^c_{kl})}
                             - 2 e^{-\beta_{kl}(r-r^c_{kl})} \},
\end{eqnarray}
where $k$ and $l$ specify the two species of atoms, say {\it host} and 
{\it guest}. 
Host and guest atoms are denoted by $A$ and $B$, respectively. 
A cluster is formed by $N_A$ host atoms and $N_B$ guest(or solute) 
atoms, where total number of atoms is $N=N_A+N_B$.
For simplicity we choose $\beta_{AA}=\beta_{BB}=\beta_{AB}=1.3588[A^{-1}]$, 
$\epsilon\equiv\epsilon_{AA}=\epsilon_{BB}=0.3429[eV]$ 
and $r^c_{AA}=r^c_{BB}=r^c_{AB}=2.866[A]$. 
Those values are suitable for copper\cite{Giri}.
The only free parameter is $\alpha=\epsilon_{AB}/\epsilon_{AA}$. 
Because the heat of solution $\Delta H$ is given by 
$\Delta H=z(1-\alpha)\epsilon$ where $z$ is a coordination number,  
our choice for $\alpha$($\alpha=1.1$), provides a negative 
heat of solution for the binary system\cite{notice}.
For a realistic binary system 
the relations, $\epsilon_{AA}=\epsilon_{BB}=\epsilon_{AB}$ 
and $r^c_{AA}=r^c_{BB}=r^c_{AB}$ 
do not hold. 
Although we may oversimplify the model system, 
we believe our model is suitable to investigate how the effect 
of the heat of solution controls SA.  \par
In our model we neglect the presence of supporting film which may be 
important as a heat reservoir. 
However, the released binding energy is transfered to a substrate very
slowly, since the coupling between a substrate and a cluster 
is considerably weak.
Weak coupling between a substrate and a cluster is 
due to the frequency mismatch between atoms in a substrate and a 
cluster(See appendix).
\subsection{Plausible initial conditions}
It is very hard to simulate the realistic condition of YM's original
experiment.  The solute atoms which are successively deposited
onto the host cluster form new bondings and release the bonding
energy as excess kinetic energy. 
This process makes the cluster so hot
that some cluster atoms suddenly evaporate, because the total 
energy is conserved in our simulation. 
We, therefore, employ an initial condition
in which the solute atoms are bounded {\it stably} with the host 
cluster at its surface as shown in Fig.2 in order to remove other 
initial conditions in which surface atoms are suddenly evaporated. 
In addition, 
as will be discussed later in detail, it is important to note that 
the temperature, which is identified with the average kinetic energy 
of the cluster, is one of the key parameters which controls SA. 
\begin{figure}
\begin{center}
\epsfysize=4cm
\epsfbox{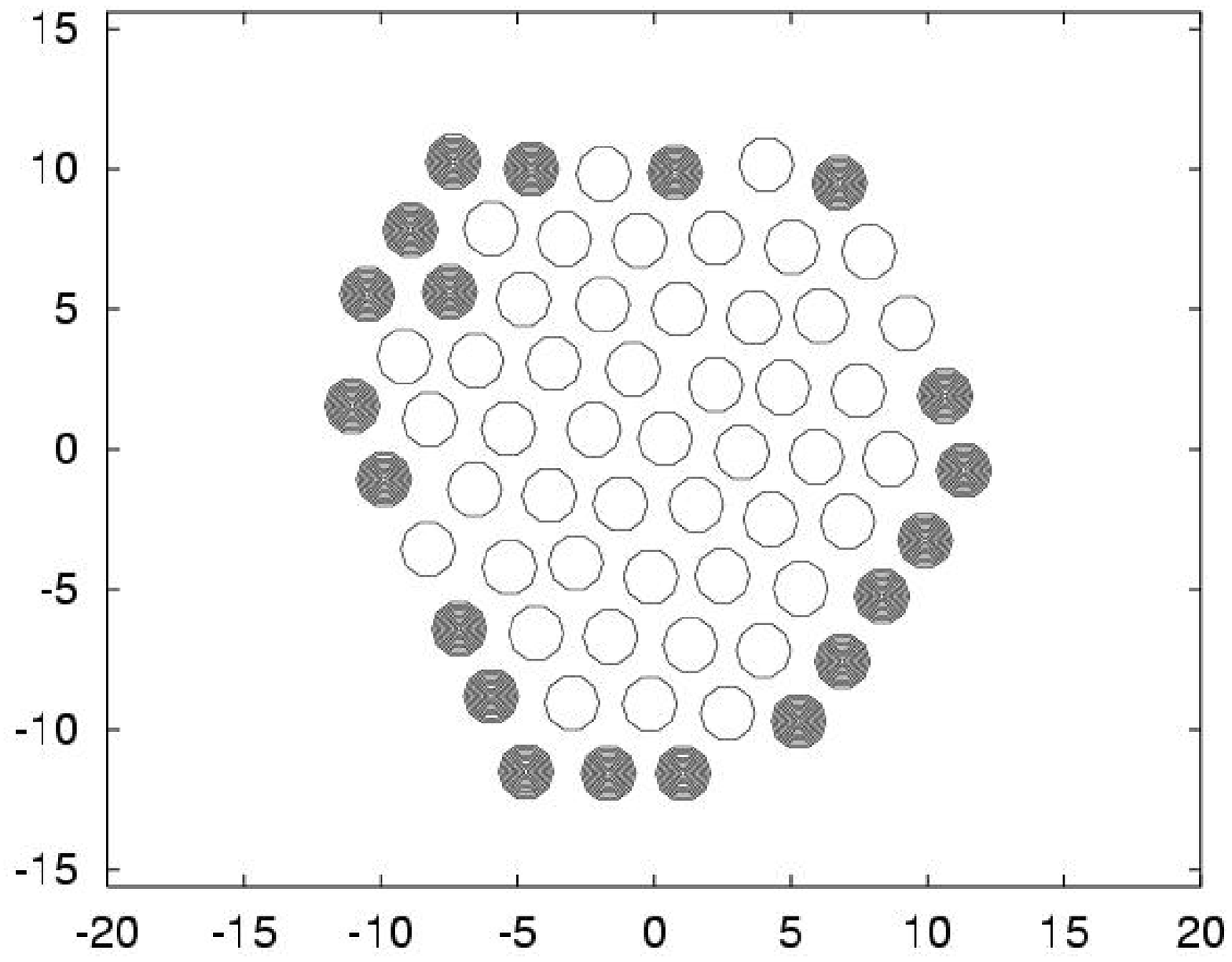}
\end{center}
{\bf Fig.2} A typical initial atomic configuration 
for isoenergetic MD simulation.
White and black circles denote host(A) and guest(B) atoms, respectively.
A cluster consists of 47 host atoms and 20 guest atoms,$(N_A,N_B)=(47,20)$.
(The radius of each circle is given by an arbitrary unit.)
\end{figure}
Thus it is desirable to keep the initial temperature of the cluster
as stationary as possible.  
Arbitrarily chosen initial configurations are {\it unstable} 
in the sense that the surface atoms do not form saturated bondings
with the inner atoms and the configuration is not located in a
sufficiently deep local minimum of the potential energy.
As a result, in time evolution of MD run,  
formation of new bonds between atoms leads to a rapid
and considerable temperature rise.
In such cases the initial temperature is not well controlled.
Indeed, we have observed, in preliminary simulations, that 
atoms attached to the surface of a cluster are in general
promptly absorbed into the first layer of a host cluster 
and release the kinetic energy in an uncontrollable manner.
These are the practical reasons why we choose a stable configuration
as is shown in Fig.2 with which the kinetic energy keeps
a stationary value over a sufficiently long time scale in which 
the initial temperature is well defined.\par
Selection of a stable initial configuration is also justified 
from the experimental viewpoint. Using nm-sized amorphous Sb(a-Sb) 
clusters, each of which was attached to a Au cluster, YM also 
confirmed the presence of SA\cite{au-sb}. 
In their experiments these binary clusters were 
gradually heated from $96K$ to $290K$. Then, dissolution of Au 
into a-Sb clusters set in around $200K$ and Sb-Au alloyed 
clusters were produced in the time scale much less than 1sec. 
Unlike the original experiment of $(Au-Cu)$ cluster by YM, the two sorts of
clusters bonded firmly with each other at their interface
and could be identified with our stable initial configuration. \par
Stable initial configuration for MD is prepared by the simulated
annealing method: prepare a homogeneous cluster composed of 
the A-atoms alone and start its time evolution, 
then the system wanders over various 
local minima of the potential energy. 
At the step when the configuration 
falls into a sufficiently deep local minima, we choose it as the initial 
condition and start the simulation of SA after assigning an appropriate number
of atoms at the surface of the cluster to the solute atoms i.e., B-atoms.
After allocating initial random velocity to each atom, 
the trajectories were computed by the velocity form of the Verlet algorithm, 
where time step used was $5\times 10^{-15} sec$ \cite{Verlet}.
Time evolution of $10^{-6}sec$ was run in every case.
\section{Tools for quantification: How to observe atomic fluctuation
and rearrangement during SA}
In this section we give some examples of numerical results 
obtained from microcanonical MD simulation. 
Our main purpose here is to introduce some useful quantities and  
to outline how SA proceeds by them.
\subsection{Useful tools } 
Prior to discuss numerical results of size, temperature and 
$\alpha$ dependency, we introduce some quantities characterizing the 
atomic motion in a cluster. 
At first, we define a quantity measuring to what extent two 
different types of atoms are mixed in a cluster.
To do this, we introduce the number of neighboring host atoms per 
a solute atoms, say $n_B(t)$. $n_B(t)$ is defined as,
\begin{eqnarray}
  n_B(t)=\frac{1}{N_B}\sum_{i=1}^{N_B} N_{A(i)}(t),
\end{eqnarray}
where $N_{A(i)}$ is the number of 
A-atoms which occupy the nearest neighbor sites  
around the i-th B-atom at time $t$.
In case that two types of atoms are homogeneously mixed, 
a simple mean field consideration yields that 
$n_B(t)$ should be $n_B^H=z(1-r)$, where $z$ and  $r$ denote 
coordination number and fraction of solute atoms respect to 
total number of atoms in a cluster. \par
During the SA process each atom in a cluster vibrates near a stable 
position and sometimes jumps to another neighboring position.
The accumulation of the latter process results in the mixing
of the solute atoms into the host atoms.
Accordingly the dynamics of alloying process has at least three
different time scales: the shortest one is due to the rapid 
{\it fluctuations} around the site, which
is comparable to the inverse of Debye 
frequency ( $\sim 10^{-1}$[ps]).  
The second one, which is much longer than the first one,
is the time scale of {\it rearrangement} of their neighboring atoms 
($\sim 10-100$ [ps]). 
The longest one characterizes the time scale of the alloying process
which is the relaxation time from the initial nonequilibrium
configuration to an equilibrium one, which
is shorter than 10msec in YM's experiments.  
Since the third time scale can be observed in terms of the variation of
$n_B(t)$, we introduce alternative quantities characterizing atomic 
motion observed during the shorter two time scales.
To quantify fluctuating and rearranging properties of atoms,  
we introduce the atomic Lindemann index $\delta(i)$ and
the frequency of recombination of the neighboring atoms.
The rapid vibration of atoms around each site 
is manifested in the so-called Lindemann index, 
which is expressed by 
the root mean square(rms) deviation
of atomic separation between neighboring atoms\cite{Berry1}. 
We define the nearest neighbor Lindemann index(NNL) for individual atoms, 
$\delta_i(t)$, as follows; 
\begin{equation}
\delta_i(t)=\frac{1}{ \langle N^{(i)} \rangle }\sum_{{\scriptstyle j \in n.n.} 
\atop{\scriptstyle 
of\ i-th\ atom}} \frac{\sqrt{\langle R_{ij}^2\rangle_t-\langle 
R_{ij}\rangle_t^2}}{\langle R_{ij}\rangle_t},
\end{equation}
where $R_{ij}$ denotes the distance between $i-th$ and $j-th$ atoms,  
$\langle N^{(i)} \rangle_t$ is  time-averaged 
number of the nearest neighbor atoms of $i-th$ atom.
Note that  $\langle {\cal F}\rangle$ is 
time average of the arbitrary quantity ${\cal F}$, given by, 
\begin{equation}
\langle {\cal F}\rangle_t=\frac{1}{t_{av}}\int_t^{t+t_{av}}{\cal F}(\tau)d\tau.
\end{equation}
The averaging time $t_{av}$ is fixed to be $2$ ns.
%
On the other hand, the frequency of recombination of the neighboring atoms
is estimated by the distance index\cite{Sawada}.
Distance index is derived from a adjacency matrix, say  ${\sf M}$,  
which is $N \times N$ symmetric matrix whose elements 
${\sf M}_{ij}=1$ for $\mid r_{ij}\mid < r_c$ and zero otherwise, where 
$r_c=1.34r_{AA}^c$.
Distance index $d_i(t)$ of the i-th atom is, then  defined as
\begin{equation}d_i(t)=\sqrt{\sum_{j=1}^N \mid  {\sf M}_{ij}(t+\Delta t)
-{\sf M}_{ij}(t) \mid ^2}
\end{equation}  
Supposing that atoms move from one site to another site frequently in
a cluster, then the occurence of atomic rearrangements should be 
manifested by the variation of the distance index $d_i(t)$. 
Time interval, $\Delta t$, must be short enough to resolve 
the single event of atomic rearrangement. 
In our numerical analysis we set $\Delta t=10ps$.
\par
\subsection{Some examples} 
Before illustrating systematic results of our simulation,
we show a typical example of time evolution of the alloying
process observed in our numerical simulation for the
$A_{47}B_{20}$ cluster (denoting the cluster of $(N_A,N_B)=(47,20)$)
and $\alpha=1.1$.
In Fig.3 we show the time evolution of $n_B(t)$, i.e., the number of 
neighboring atoms of different species per a solute atom.
$n_B(t)$ increases monotonically from the initial value $n_B(0)\sim 2$, 
which means that the solute B-atom initially on the surface of the cluster 
forms two bonds with the host A-atoms in the inner shells.
Finally $n_B(t)$ reaches to the level of $n_B^H$ corresponding 
to the homogeneously mixed state, which is indicated by the dotted line. 
\begin{figure}
\begin{center}
\epsfxsize=7cm
\epsfbox{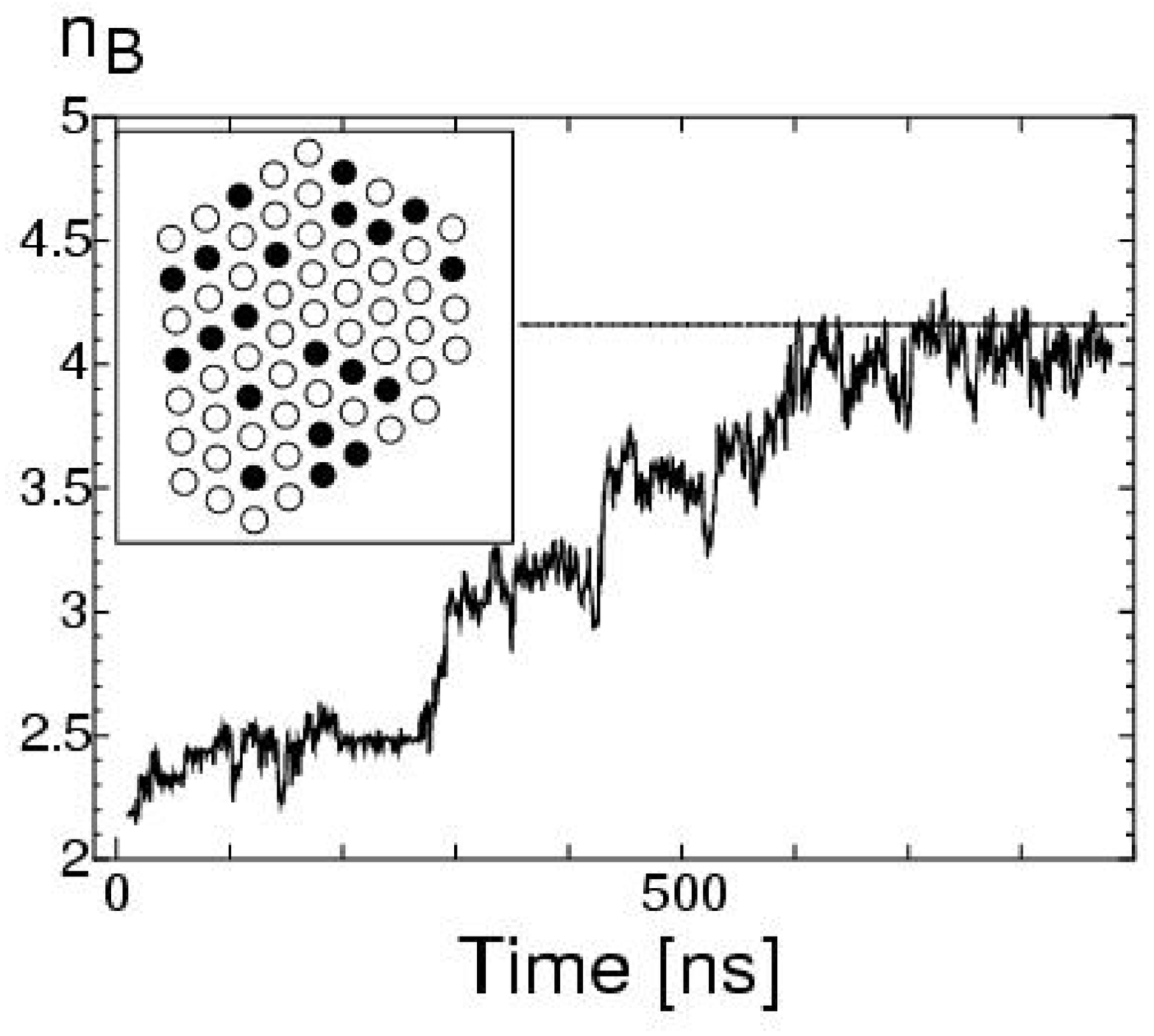}
\end{center}
{\bf Fig.3} A typical time evolution of $n_B(t)$ 
for isoenergetic MD sumulation. The final configuration
of atoms in the cluster is also inserted.
Initial temperature is about $510$[K].(See Fig.4.)
\end{figure}
It is obvious that a homogeneously mixed configuration is almost achieved 
until $800[ns]$. The present example corresponds to the case where the initial 
temperature is the lowest among our data exhibiting SA within 1000[ns].
The spontaneous increase of the $n_B(t)$ implies that the system
evolves so as to decrease the potential energy of the system.
Indeed, the increment,$n_B(t)-n_B(0)$, means that the variation of 
potential energy per one B-atom is $(n_B(t)-n_B(0))z(1-\alpha)$.
The time averaged variation of the total potential energy, then, is
$\langle \Delta U(t)\rangle_t \cong N_B(n_B(t)-n_B(0))(1-\alpha)$,
if we take into account the contribution from the nearest neighboring atoms.
The decrease in the potential energy is converted into the increase 
in the kinetic energy. 
In our simulation we define the kinetic temperature by 
$T=\frac{2E_k}{k_B(2N-3)}$,
where $E_k$ is total kinetic energy of the system and $k_B$ is 
Boltzmann constant. 
Note that we eliminate the contributions from translational and
angular degrees of freedom, because we select 
initial conditions with vanishing translational and angular momentum.
Variation of the kinetic temperature of the cluster is shown in Fig.4.
\begin{figure}
\begin{center}
\epsfxsize=7cm
\epsfbox{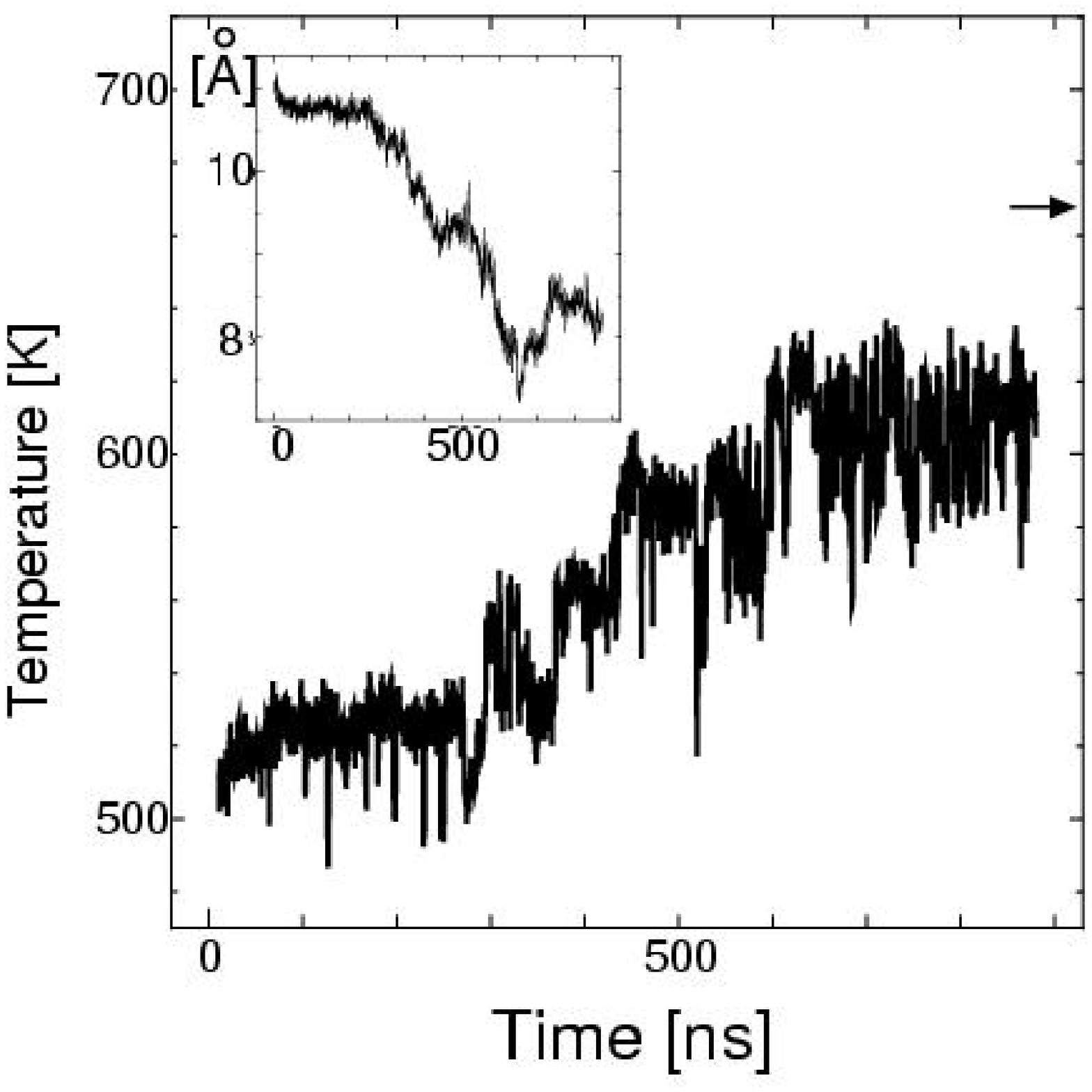}
\end{center}
{\bf Fig.4} A typical time evolution of kinetic temperature 
for isoenergetic MD simulation of $A_{47}B_{20}$ binary Morse cluster 
taken from the same data as Fig.3. Initial temperature is estimated 
as $510[K]$. Averaging of kinetic temperature was done in every $2[ns]$.
Notice that resulting temperature does not exceed estimated 
melting temperature($\sim 670[K]$)(See also Table.I).
\end{figure}
As is expected, it is clear that the variation of $n_B(t)$ is  
strongly correlated to the temperature rise. 
The final increase in the temperature, say $\Delta T$, is less than  
$\Delta T_{max} = -\frac{2}{k_B (2N-3)}\langle \Delta U(t=\infty)
\rangle_{t_{av}} \sim 200$[K], 
and the half of $\Delta T_{max}$ contributes  
to the actual temperature rise according to the virial theorem.
Indeed the virial theorem predicts $ \Delta T\sim 100 K$, 
which is consistent with Fig.4.
Although the temperature is increased up to about $610$[K] when the alloying 
process is completed, the temperature at $t=800[ns]$ 
is still sufficiently below the melting temperature of the cluster.  
The typical melting temperatures measured for various sizes of clusters 
is listed in Table I.\par
Although one cannot observe a sharp solid-liquid transition 
in a small system in a strict sense, 
we can practically locate the melting point by 
an abrupt jump in caloric curve and Lindemann index
\cite{Berry1,SIS}.
As is well-known experimentally and numerically, the melting 
point is reduced as size of cluster decreases\cite{Borel}.
It is evident that Table I also indicates the same trend.\\
\\
{\bf Table.I} The relation between cluster size and the melting 
temperature for our 2D Morse model.\\
\begin{tabular}{c|c} \hline
Number of atoms & Melting tempetature [K]\\ \hline
32 & 580 \\
67 & 670 \\
80 & 710 \\
117 & 750  \\ \hline
\end{tabular}
\\
\\
Since the melting temperature of a cluster $A_{47}B_{20}$ 
is about $670$[K], dynamics of SA process in Fig.4 
provides an evidence manifesting that the alloying process completes 
in the solid phase without the melting
of the whole cluster.\par
In Fig.5 we show the snapshots of the atomic configurations 
corresponding to SA process depicted in Fig.3 and 4.
\begin{figure}
\epsfxsize=8.5cm
\epsfbox{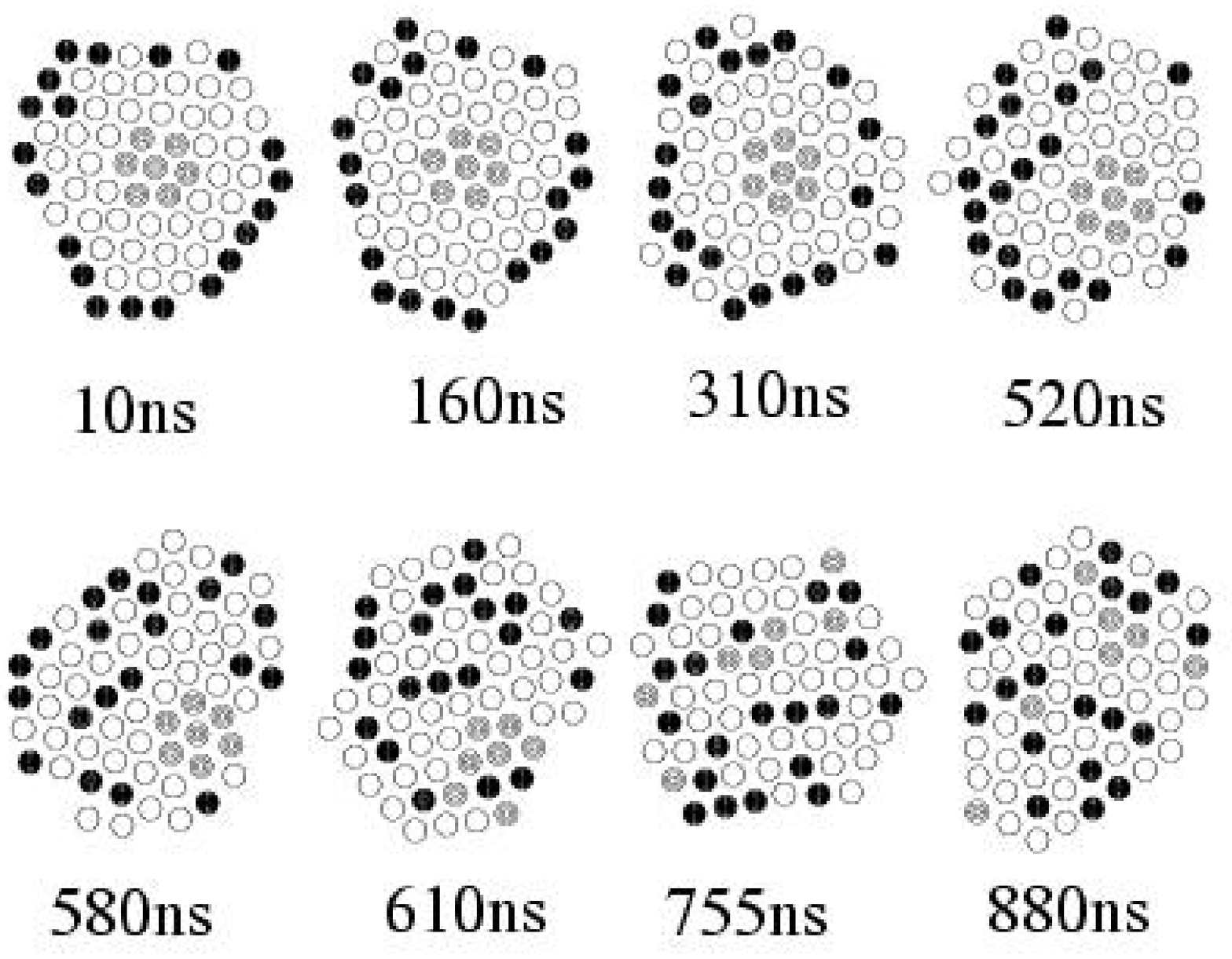}
{\bf Fig.5} Snapshots of atomic configuration of $A_{47}B_{20}$ 
in time evolution.
White and black circles denote host and guest atoms, respectively.
Shaded circles are host atoms which initially forms core of 
a cluster.
\end{figure}
The B-atoms initially deposited on the surface wanders actively {\it along} 
the surface of the cluster.
However, as clearly seen in Fig.5, they all stay on the surface 
and the movement in the radial direction of the cluster 
is almost quiescent over a significantly long time ($t<300$[nsec]). 
The length of the quiescent period depends very sensitively on the initial 
configuration.
After that the radial movement begins to be activated. 
Some atoms suddenly enter 
into the inner shell, which is 
reflected in an abrupt rise of $n_B(t)$ in Fig.3.   
Then the B-atoms enter further into inner layers and $n_B(t)$ increases
stationarily. 
During the stationary stage, the B-atoms aggregate in one side 
of the cluster as is typically seen in the atomic configurations
at $t=520-580$ [nsec].
The atoms initially forming the core of the cluster, 
which are indicated by shaded atoms in snapshots, 
are pushed out in a group and breaks up when they reach the surface.
In this way the outer B-atoms exchange their positions 
with the inner A-atoms. 
All the alloying processes observed in our microcanonical simulation 
completes according to the similar scenario.\\
\section{ Dependence of the alloying process on key parameters}
According to the conclusions by YM,
SA phenomena are dominated by three key parameters; 
magnitude of negative heat of solution, temperature
of the system and the size of a cluster. 
In this section we show systematic numerical results
exhibiting that these parameters are similarly important 
for the onset of SA reproduced in our microcanonical 
simulation. 
\subsection{A role of heat of solution as a driving force of SA}
In Sec.IV we illustrated a typical dynamical behavior of SA 
which is driven by negative heat of solution. 
In order to probe the dependence of alloying process on heat of solution
we compare three different cases, i.e., $\alpha=0.9$(positive
heat of solution), $\alpha=1.0$(vanishing heat of solution) 
and $\alpha=1.1$(negative heat of solution).\par
In Fig.6(a)(b), we show the time evolution of $n_B(t)$ for 
two values of initial temperature, $620[K]$ and $510[K]$, respectively. 
In both cases we also display the time evolution of $n_B(t)$ for 
$\alpha=0.9$, $\alpha=1.0$, and $\alpha=1.1$. \par
Some significant discrepancies in the variation of $n_B(t)$ among 
these three cases become obvious in Fig.6 (a)and (b). 
It is clear that the value of $n_B(t)$ for $\alpha=1.1$ 
shows a rapid increasing trend which is a signature of 
a faster alloying process.
Conversely, an absence of a mixing between guest and host atoms is  
manifested in a slower decrease and saturation of $n_B(t)$ for $\alpha=0.9$.
For $\alpha=1.0$, $n_B(t)$ increases very slowly, because A- and B-atoms  
are mixed to some extent. 
However, it does never reach the value, $n_B^H$, within the 
simulation time($800$[ns]).
The difference in the variation of $n_B(t)$ is a direct evidence 
indicating that the SA process is dominated by heat of solution. 
In other words, time to complete SA becomes shorter as 
initial temperature is getting higher.   
A systematic analysis to clarify the relationship between  
the alloying speed and initial temperature is pursued in Sec V.B. 
\par
\begin{figure}
\begin{center}
\epsfxsize=8cm
\epsfbox{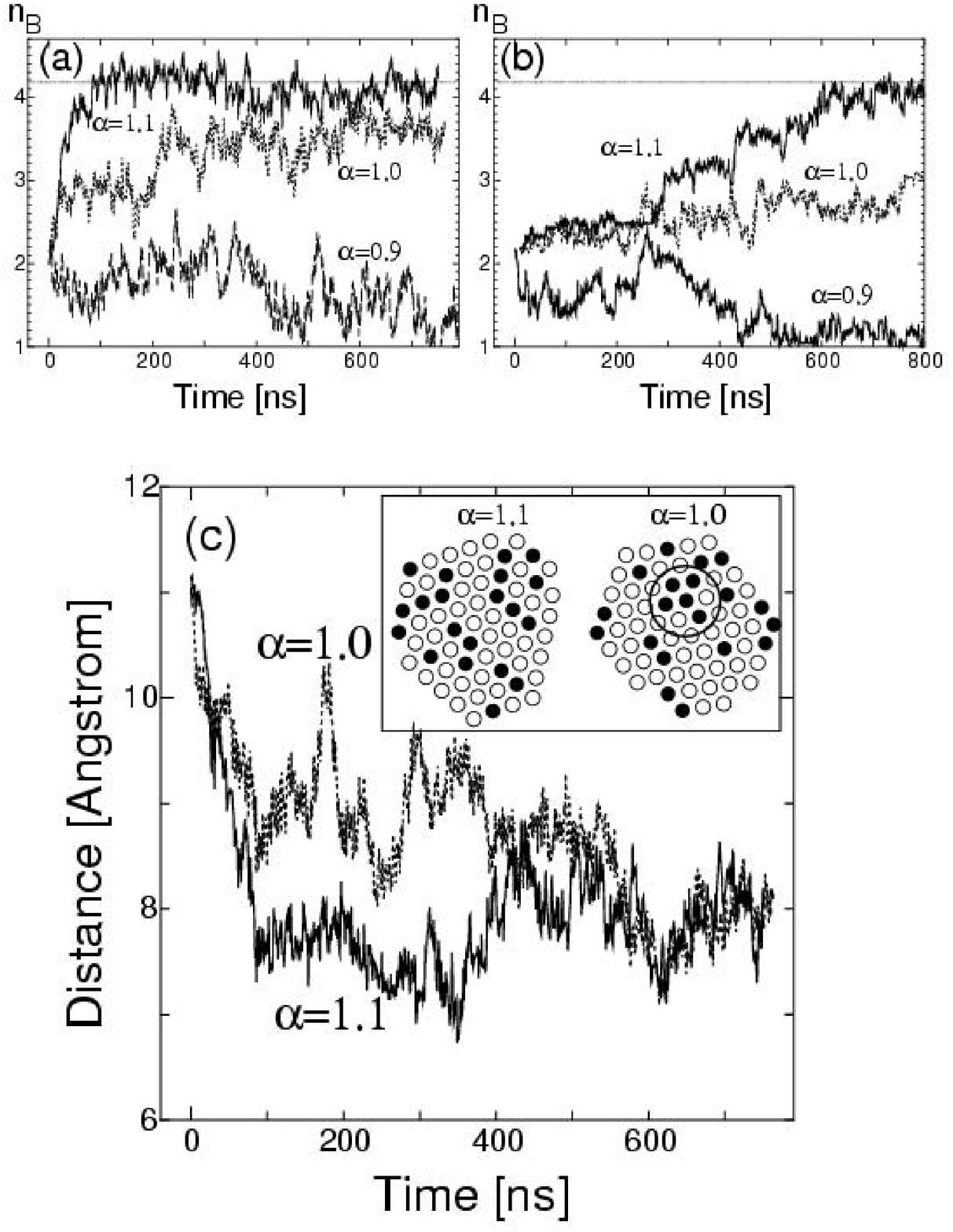}
\end{center}
{\bf Fig.6} A typical example of time evolution of $n_B(t)$, which is 
defined for host atoms(A-atoms) and guest atoms(B-atoms),   
obtained by isoenergetic MD runs with high and low 
initial temperature.((a)$610$[K](high) and (b)$510$[K](low)).
(c)The radial diffusion of atoms in a cluster is evaluated in 
terms of the time evolution of the distance between  
B-atoms and the center of mass of the cluster for the case(a). 
The final configurations of a cluster for $\alpha=1.1$ and 
$\alpha=1.0$ are also inserted. 
An aggregation of B-atoms in a cluster is also indicated 
by the circle of the inserted figure for $\alpha=1.0$. 
\end{figure}
In addition to the main features mentioned above, 
it is important to note the following numerical results.\\
(1)For the relatively high temperature, the mixing behavior
takes place even for the atomic species with null heat 
of solution ($\alpha=1.0$).
As shown in Fig.6(c), most of the B-atoms diffuse into 
the cluster even for the case of $\alpha=1.0$.
That is, the average distance between 
the B-atoms and the center of mass reaches approximately $8.8$, which 
is very close to the value of the final configuration of 
$\alpha=1.1$, where a complete mixing is attained. \\
(2) On the contrary, $n_B(t)$ for $\alpha=1.0$ 
is always smaller than that for $\alpha=1.1$.\par
These seemingly contradictory facts are understandable by comparing 
the final atomic configurations for $\alpha=1.0$ with that for 
$\alpha=1.1$.
As shown in Fig.6.(c), the A-atoms and the B-atoms are well separated for 
$\alpha=1.1$, while the B-atoms tend to aggregate for $\alpha=1.0$. 
Aggregated configurations of the A- and the B-atoms are energetically 
neither favorable nor unfavorable for $\alpha=1.0$, 
although such an aggregation of the B-atoms are 
energetically unfavorable for $\alpha>1.0$.
Indeed, as indicated by the circle in the inserted figure of Fig.6(c), 
an aggregation of the B-atoms is easily verified.    
In short, in case of $\alpha=1.1$, the A- and the B-atoms 
mix so as to decrease the potential energy and 
it works as the driving force of SA. 
Consequently, the role of negative heat of solution is twofold, 
as far as our microcanonical simulation is concerned.
First, it results in a driving force which promotes
the alloying of two species of atoms. 
Secondly, the decrement of potential energy 
in the alloying process is converted into the kinetic energy, 
which heats up the cluster and accelerates SA. 
The latter effect is persued again in the next section.
However, these two facts indicate that the 
diffusion of the B-atoms into the cluster from the surface 
is by no means prohibited even in case of 
$\alpha=1.0$, where the potential energy gain due to 
the mixing is zero. 
It should be noted that such a diffusion process with 
is still much quicker than the diffusion into the bulk media. 
We numerically confirmed that no significant diffusion occurs in
the bulk 2D medium within the time scale of Fig.6 at the same 
temperature. \par
In summary, heat of solution is a key parameter of 
SA in the sense that positive heat of solution prohibits SA, 
while negative heat of solution
remarkably accelerates SA. 
However, a rapid mixing of the two species of atoms occurs 
even in case of null heat of solution. 
This fact demonstrates that the rapid diffusion process is 
a generic feature of microclusters. 
The present result is consistent with YM's experiments. 
Indeed, YM reported that SA occurs even in the combinations 
of atomic species with very small magnitude(almost null) 
of positive heat of solution, when the size 
of cluster is sufficiently small\cite{pheat}.
\subsection{Temperature dependence and size dependence}
In YM's experiments, the time needed for 
the SA process to complete depends sensitively on the temperature.
Some systematic results for the initial temperature dependency of SA 
is presented here. 
Because kinetic temperature is not constant during time evolution 
in a microcanonical dynamics, we regard the initial kinetic temperature
as the parameter characterizing the temperature of the system. 
The temperature dependence of the alloying
time is examined for the four sizes of clusters 
$A_{23}B_9$, $A_{47}B_{20}$,
$A_{56}B_{24}$, and $A_{140}B_{60}$.  
For each of them, we prepare initial conditions 
corresponding to various values of initial kinetic temperature.
The values of initial kinetic temperature cover a 
wide range from well below to just below the melting temperature.
The value of the paramter $\alpha$ is fixed ($\alpha=1.1$) 
for all sizes of clusters.
Note that the ratio $r=\frac{N_B}{N_A+N_B}$
is chosen to be the same value, $r=0.3$, for the four sizes of clusters.
This is because the bonding number of homogeneous mixing $n_B^H$ 
should be fixed to be a common value irrespective of the cluster size. 
Moreover, the possible temperature 
rise $\Delta T =\frac{1}{2k_B}[n_B^H-n_B(0)]r(\alpha-1)$
in the SA process are controlled to be common in the 
four sizes.\par
The {\it alloying time} should be defined as the time 
required to attain the homogeneous mixing, 
which corresponds to the time when $n_B(t)$ reaches 
the value for homogeneous  mixing i.e.,$n_B^H=4.2$. 
However, it takes extremely long time to complete SA 
especially for larger clusters, and it is difficult 
to get reliable data. 
We, therefore, define the alloying time $\tau_{alloy}$ as the time when 
$n_B(t)$ reaches $3.0$, which is almost the average of the initial 
value of $n_B$ and $n_B^H$. 
The semi-log plot of the inverse initial temperature
versus the alloying time is depicted in Fig.7 
for the four clusters.
\begin{figure}
\begin{center}
\epsfxsize=7cm
\epsfbox{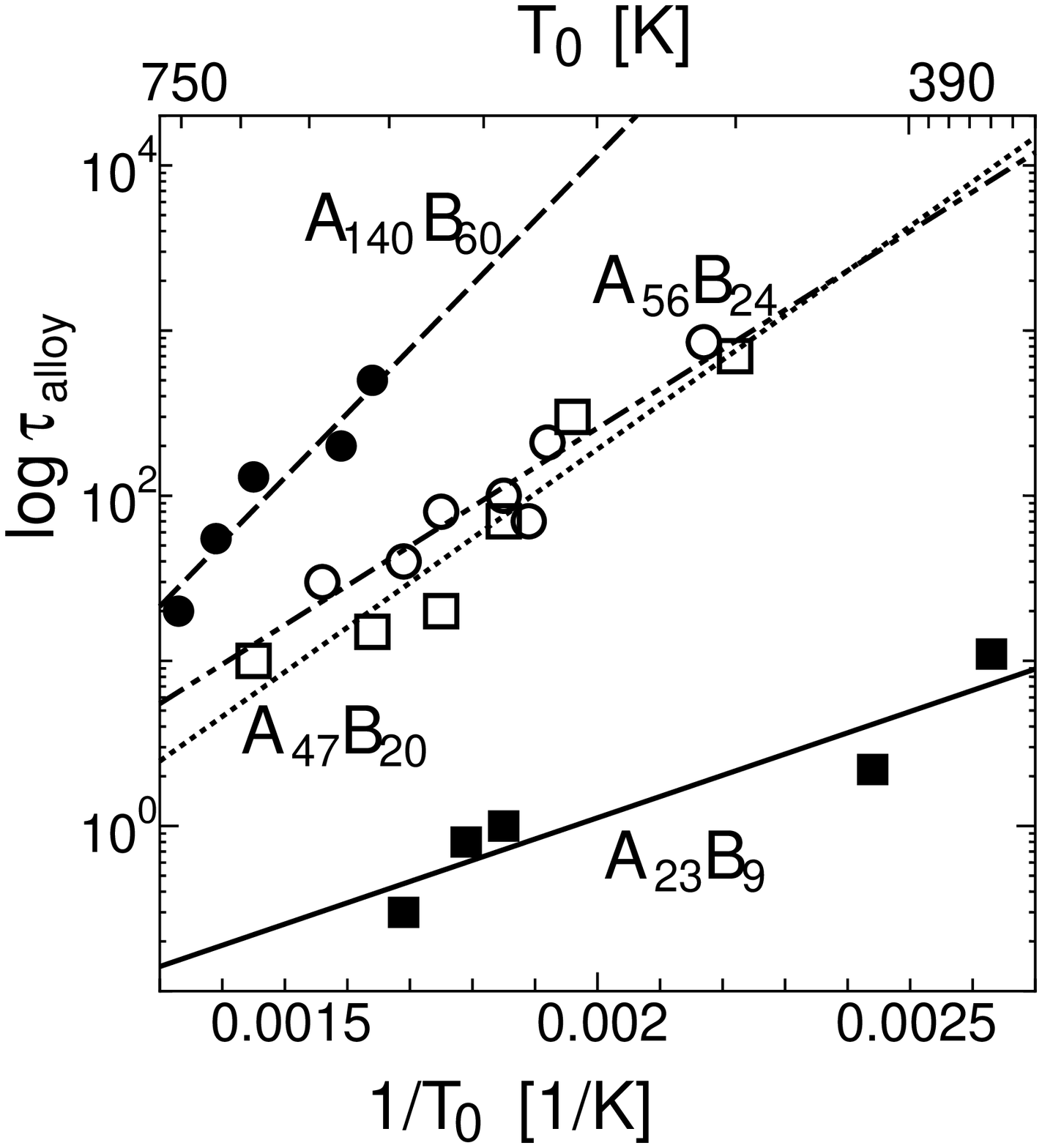}
\end{center}
{\bf Fig.7} 
The dependence of the alloying time $\tau_{alloy}$
on the initial temperature $T_0$ for 2D Morse 
binary cluster $A_{23}B_9$(black square), $A_{47}B_{20}$(white square), 
$A_{56}B_{24}$(white circle), and $A_{140}B_{60}$(black circle).
\end{figure}
Because SA is an outcome of the diffusion of solute atoms 
into the cluster, it is reasonable to expect that 
the alloying rate also obeys the Arrhenius-like 
law of the diffusion coefficient.
Since the time evolution pattern $n_B(t)$ largely fluctuates from
sample to sample, the alloying time defined above
is also accompanied with considerable sample-dependent fluctuations.
However, as shown in Fig.7 the semi-log plots for these samples
are almost on lines. 
Aftrer all, the dependence of the alloying time on the initial 
temperature obeys an Arrhenius-like law:
\begin{eqnarray}
\tau_{alloy} = \tau_0 \exp (\frac{T_{alloy}}{T_0}),
\end{eqnarray}
where  $T_{alloy}$ can be interpreted as the 'activation 
energy' divided by Boltzmann constant and $\tau_0$ is the inverse of 
the frequency factor. 
It is evident that the activation energy, which is the slope 
of each line, increases with the size of the cluster.
The present result is the manifestation of the size effect which
was observed in YM's experiments and is regarded as
the characteristic effect peculiar to clusters.
The activation energy $T_{alloy}$ and the inverse of the 
frequency factor $\tau_0$ evaluated for the four sizes of 
clusters are summarized in Table II. \\
\\
{\bf Table II} Numerically estimated $\tau_0$ and $T_{alloy}$ for 
various sizes of binary clusters,i.e. 
$A_{140}B_{60}$, $A_{56}B_{24}$, $A_{47}B_{20}$ and $A_{23}B_{9}$.\\
\begin{tabular}{l|c|c}\hline
Cluster Size  & $\tau_0$ [psec] & $T_{alloy}$ [K]\\ \hline
$A_{140}B_{60}$ & $0.02$  & 8600 \\
$A_{56}B_{24}$ & $2.7$ & 5400 \\
$A_{47}B_{20}$ & $0.91$ & 6000 \\
$A_{23}B_{9}$ & $2.7$ & 3000 \\ \hline
\end{tabular}
\\
\\
By an extrapolation with these values we roughly estimate
the alloying time at room temperature. 
The estimated alloying times are  $5.4\times 10^{-10}$[sec], 
$7.3\times1.0^{-4}$[sec], $4.4\times1.0^{-4}$[sec] and 
$1.3$[sec] for $A_{23}B_{9}$, 
$A_{47}B_{20}$, $A_{56}B_{24}$ and $A_{140}B_{60}$ at $300K$, 
respectively. 
The resulting values are short enough to be consistent with 
YM's experimental observation.\par
In addition, it is interesting to note the fact that activation energy 
$T_{alloy}$ has similar values for $A_{56}B_{24}$ and $A_{47}B_{20}$, 
although the size of the former is larger than that of the latter.
The apparent contradiction is understandable, if we 
assume that these clusters consist of the same number of shells. 
(The precise definition of the shell is given in Sec.VI.A.)
One can easily verify that the clusters, $A_{23}B_{9}$, $A_{47}B_{20}$,
$A_{56}B_{24}$ and $A_{140}B_{60}$ are composed of 2,4,4, and 7 shells, 
respectively. 
Thus, the cluster size represented by the number of shells is 
a relevant quantity to determine the activation energy given in Table.II.
\section{An activity of a cluster surface and the mechanism of SA}
\subsection{A shell dependent activity of atoms}
In this section we probe how actively individual atoms 
composing the cluste move during SA. 
To evaluate activity of atomic motion in a cluster provides 
important clues to elucidate the mechanism of SA process.  
In particular, we pay our attention to how 
dynamical activities represented by {\it fluctuation} 
and {\it rearrangement} of atoms depends on the distance from
the center of cluster. 
To do this, it is convenient to divide the cluster into shells. 
According to the distance of a target atom from the center atom, 
which is defined as the atom 
closest to the center of mass of the cluster, we allocate the 
shell index to each atom. 
Since a single cluster has a hexagonal structure, 
it is possible to introduce {\it magic number} where a cluster forms 
a geometrically packed configuration. 
For instance, a cluster which contains 7, 19, and 37 atoms are magic number 
clusters which consist of 1, 2, and  3 closed shells, respectively.  
A cluster containing 67 atoms, which is shown in Fig.3, is divided 
into 4 shells, say $m=4$. 
The shell index number $m$ is assigned in order of the distance 
from the center of mass. The center of mass atom is allocated 
to the zero-th shell.
In Fig.8 the frequency distribution of the distance of atoms from 
the center of mass is depicted.  
This is a typical example obtained from a single isoenergetic MD 
run of a $A_{47}B_{20}$ cluster. 
\begin{figure}
\begin{center}
\epsfxsize=6cm
\epsfbox{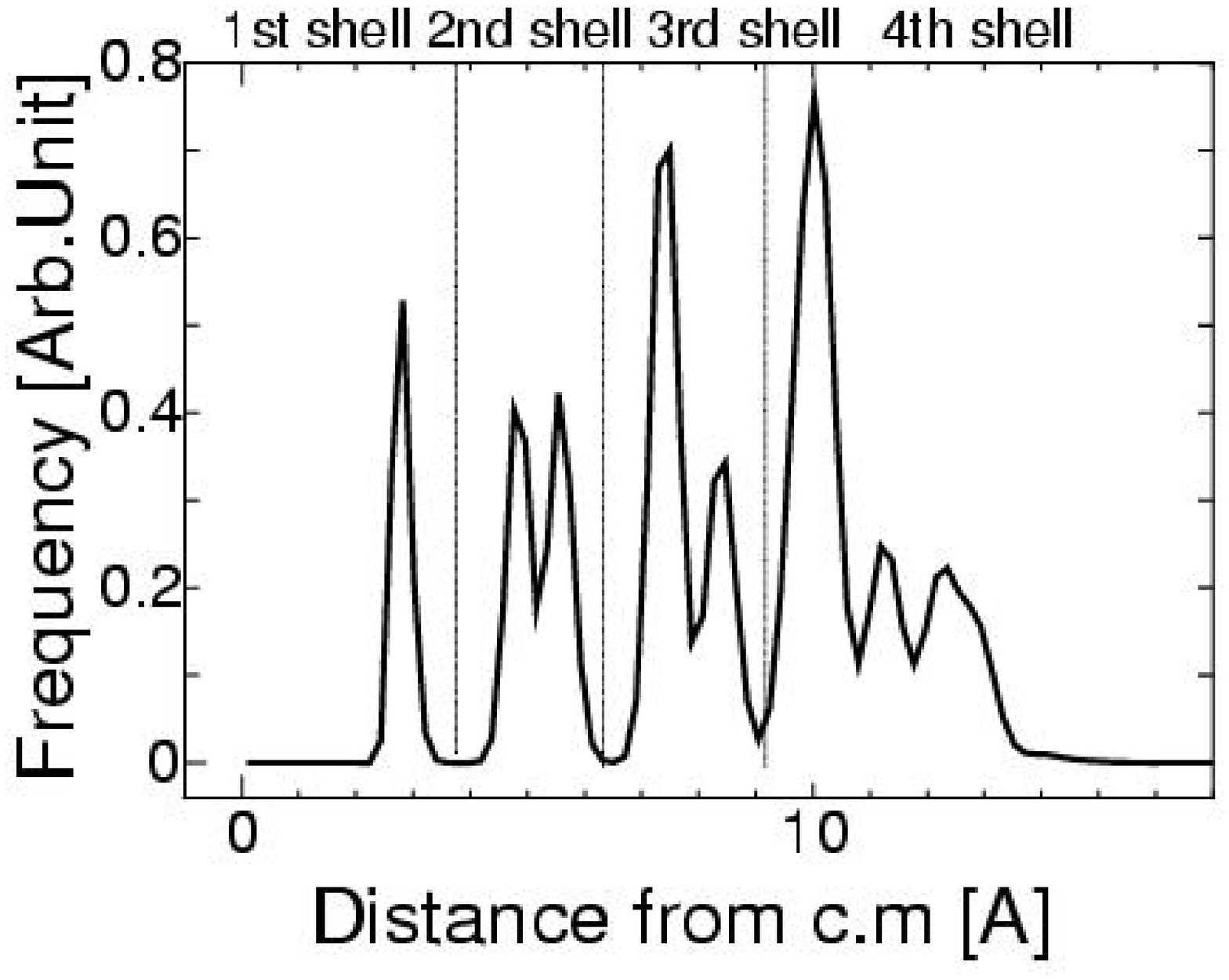}
\end{center}
{\bf Fig.8} 
A frequency distribution of a distance between the center atom and 
other atoms in a 2D Morse binary cluster $A_{47}B_{20}$. 
The dotted line denotes critical distances used to 
divide a cluster into shells. 
\end{figure}
Since the distribution has distinct minima in Fig.8, it is possible to 
divide a cluster into shells without any redundancies.
In the followings we argue about the fluctuation and the rearrangement of  
atoms in an individual shell.\par
A fluctuation property of the separation between nearby two 
atoms has been used as an indicator to locate the 
melting point\cite{Berry1}.
In fact, the location of the melting point determined by 
rms bond length fluctuation(Lindemann index) almost
coincides with the one given by the caloric curve\cite{Berry1,Kaelberer}. 
In order to get further detailed information on atomic fluctuation
we use the nearest neighbor Lindemann index(NNL) defined for each shell.
The NNL index of the $k-th$ shell,  
$\Delta_k(t)$, is defined by averaging the NNL index for individual
atoms, say $\delta_i(t)$, belonging to the same shell;
\begin{equation}
\Delta_k(t)=\frac{1}{\langle N_S^{(k)} \rangle}\sum_{\scriptstyle{i\in k-th\ 
shell}}\delta_i(t),
\end{equation}
where $\langle N_S^{(k)} \rangle$ is time average of the total 
number of atoms contained in the $k-th$ shell over $t_{av}$. \par
One can expect that some shell-dependent dynamical activities
are captured by $\Delta_k(t)$. 
In Fig.9 we depict the time evolution of $\Delta_k(t)$ 
for a single run of SA process.
\begin{figure}
\begin{center}
\epsfxsize=8.8cm
\epsfbox{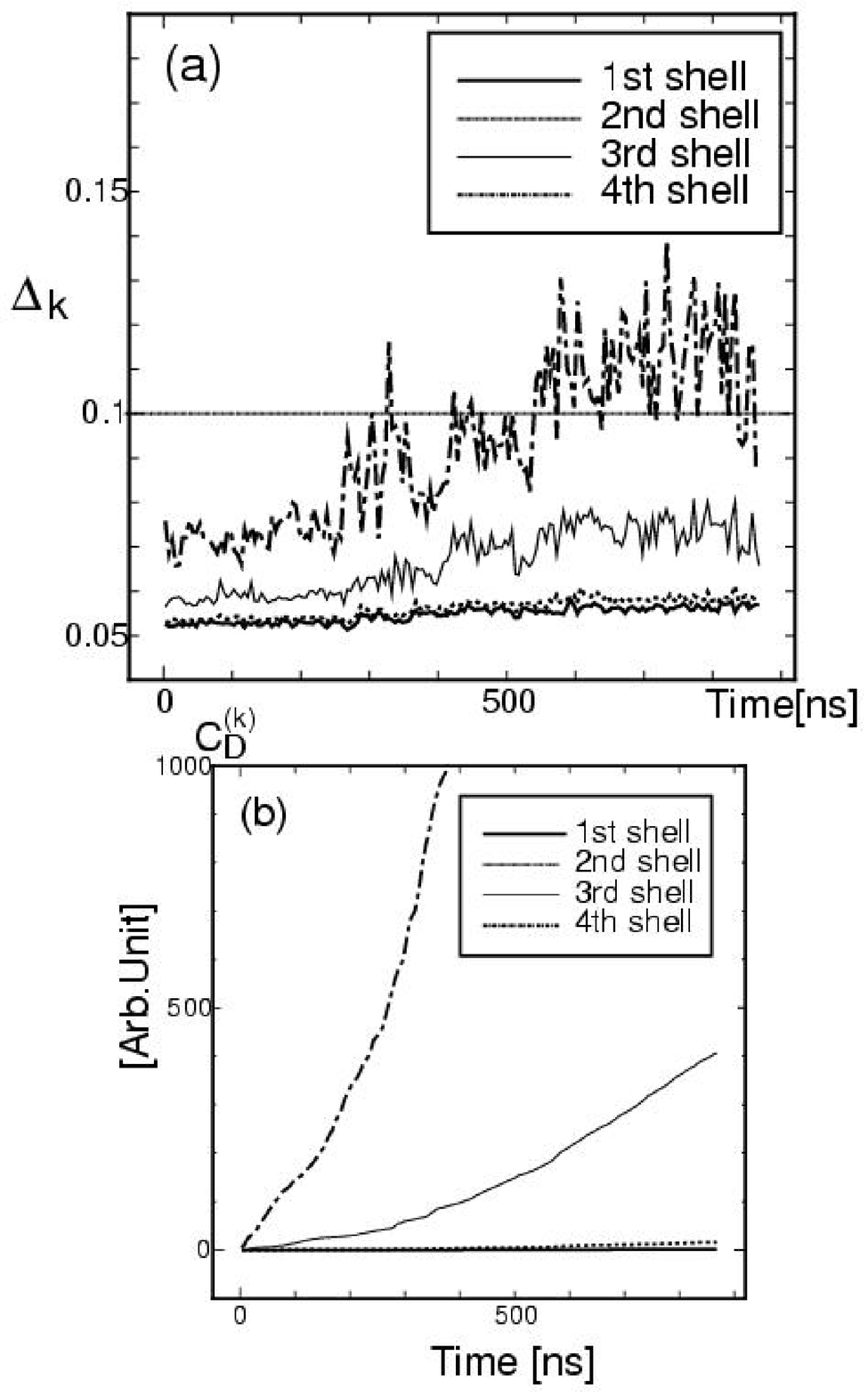}
\end{center}
{\bf Fig.9} 
(a)Time evolution of Lindemann index for the 1st, 2nd, 3rd and 4th shell 
obtained from data of 2D Morse binary cluster $A_{47}B_{20}$.
(b)Time evolution of the cumulated distance index $C_D^{(k)}$ for 
$(k=1-4)$. Initial temperature is about $510$[K].(See Fig.4.)
\end{figure}
The bond fluctuation is much enhanced in the shells near the surface
of a cluster even if its temperature is substantially below the melting 
point, which is about $680$[K]. 
It exceeds the Lindemann criterion for melting i.e., $\Delta_k \sim 0.1$,
denoted by the broken line in Fig.9.
The time evolution of $\Delta_4$ in Fig.9(a) strongly
suggests that the cluster surface is in a melting state.
Indeed the {\it surface melting} is observed in Pb cluster 
below melting temperature\cite{SM}. 
Judging from the fact that the Lindemann index of the inner 
shells are less than $0.1$, the inner shells of the 
cluster is in a solid phase in the sense of Lindemann's criterion.\par
The dependence of the Lindemann index for the 3rd and the 4th shell upon 
the initial temperature is shown in Fig.10.
\begin{figure}
\begin{center}
\epsfxsize=6cm
\epsfbox{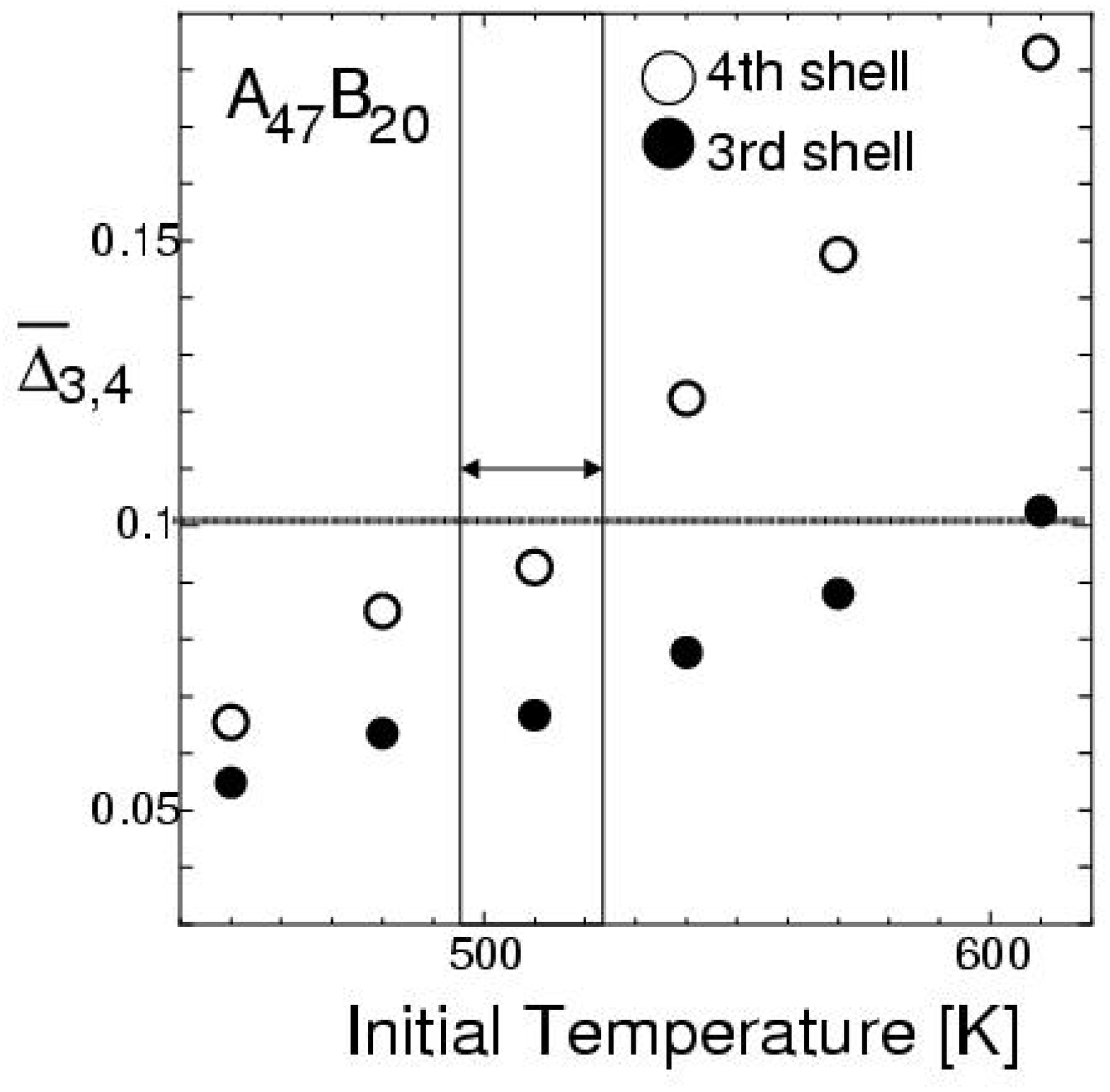}
\end{center}
{\bf Fig.10} 
The relationship between the initial temperature and 
the averaged atomic fluctuation for the 3rd and the 4th shell  
in a 2D Morse binary cluster $A_{47}B_{20}$. 
The averaged atomic fluctuation is given by $\bar{\Delta}_k=
\frac{1}{\tau_{alloy}}\int_0^{\tau_{alloy}}
dt \Delta_k(t)$. 
The black and white circles denote the Lindemann index for the 
3rd and the 4th shell, respectively. 
The surface atoms on the 4th shell begin to melt near the region 
denoted by the arrow.
\end{figure}
While the outer most shell begins to melt about $510$[K], 
SA completes within $800$[ns] at the same temperature 
as shown in  Fig.3 and 4. 
Consequently, it is plausible to say that the presence 
of the active surface atoms is necessary condition to 
attain SA within 1$\mu sec$.\par
The Lindemann index measures the fluctuation occurring on 
the time scale of atomic vibration, which is the shortest time scale 
of the system.  
An enhancement of fluctuation on the surface  
also increases the chance for the surface atoms to jump to another
vacant sites on the surface.
Such a rearrangement process is a rare event which occurs
in time scale much longer than the vibration. 
On the surface, however, the atomic rearrangement   
is also enhanced, because the surface is filled 
with {\it defects} or {\it vacancies} to which 
surface atoms are allowed to jump.  
Ease of jump on the surface enhances the diffusion constant 
along the surface.
For a demonstration of a pecular motion of surface atoms, 
it is interesting to show how the activity of atomic 
rearranging process depends on the shells.  
In order to quantify it in each shell we define the cumulated 
distance index for the $k-$th shell,
\begin{equation}
C_D^{(k)}(t)=\int_0^{t}d\tau D_k (\tau),
\end{equation}
where
\begin{equation}
D_k(t)=\frac{1}{N_s(k)}\sum_{\scriptstyle{i\in k-th\ shell}}d_i(t).
\end{equation}
where $d_i(t)$ is the distance index introduced by
eq.(6), and  $C_D^{(k)}(t)$ is the accumulated number of 
rearranging events which occur near the atoms belonging to $k-$th shell.
As easily verified in Fig.9(b), atomic rearrangement is much more frequent
in the surface shell, and almost all the rearranging events
occur in the shells on or close to the surface.
\subsection{Surface melting and the rapid radial diffusion}
It is possible to demostrate that the surface activity 
is responsible for the SA process in an alternative way.
Suppose that we could suppress the activity of surface atoms, 
for example, by embedding the alloying cluster in a bulk medium, 
then the rapid alloying process would be inhibited because the 
diffusion in the radial direction of the cluster is nothing 
more than the diffusion in a bulk medium. 
The active rearrangement of the surface atoms
is a necessary condition for the rapid alloying to be realized.
The frequency of the rearrangements per unit time 
is represented by $D_k(t)$, which increases as
the time elapses due to the temperature rise.
Its average over a single alloying process, say $\overline{D_k}$, 
obeys Arrhenius-type law respect to initial temperature $T_0$;
\begin{equation}
\overline{D_k} \propto \exp [-\frac{T_R}{T_0}], 
\end{equation}
where 
\begin{equation}
\overline{D_k}=\frac{1}{\tau_{alloy}}\int_0^{\tau_{alloy}} dt' D_k(t').
\end{equation}
This nice property enables us to introduce 
the activation energy $T_R$ of the atomic rearrangement. 
For the cluster $A_{56}B_{24}$,
we obtain $T_R \sim 3700$[K] for the surface shell. 
On the other hand, as shown in Table.II, 
the activation energy of alloying, $T_{alloy}$, is 
about $5400$[K]. 
These two activation energies are significantly different,  
\begin{equation}
\frac{T_{alloy}}{T_R} \sim 1.5 .
\end{equation}
It is possible to interpret $T_R$ as the activation energy 
of rearranging motion parallel to the surface, where $T_{alloy}$ 
measures the activation energy for the 
atoms to diffuse along the direction perpendicular to the surface. 
Let $P_{\parallel}$ and $P_{\perp}$ be the probabilities
for surface atoms to jump to another site parallel to and
perpendicular to the surface, respectively.
Supposing that the radial hopping is induced by the parallel
rearrangement, the relation between these  
two probabilities is expressed as,
\begin{equation}
\frac{P_{\perp}}{P_{\parallel}} = e^{-T_{RP}/T},
\end{equation}
where $T_{RP}=(T_{alloy}-T_{R})>0$. 
A plausible interpretation of Eqs.(13) and (14) is 
that the diffusion in the radial 
direction of cluster is the outcome of a surface 
rearrangement followed by a certain activation process 
characterized by the barrier height $T_{RP}$, which is roughly 
evaluated as $\frac{1}{2}T_{R}$.
The present interpretation is also supported by a direct observation 
of trajectories of atoms during the SA process.
In Fig.11  the trajectories of all solute atoms(B-atoms) are shown for 
every 250[ns]. 
\begin{figure}
\begin{center}
\epsfxsize=9cm
\epsfbox{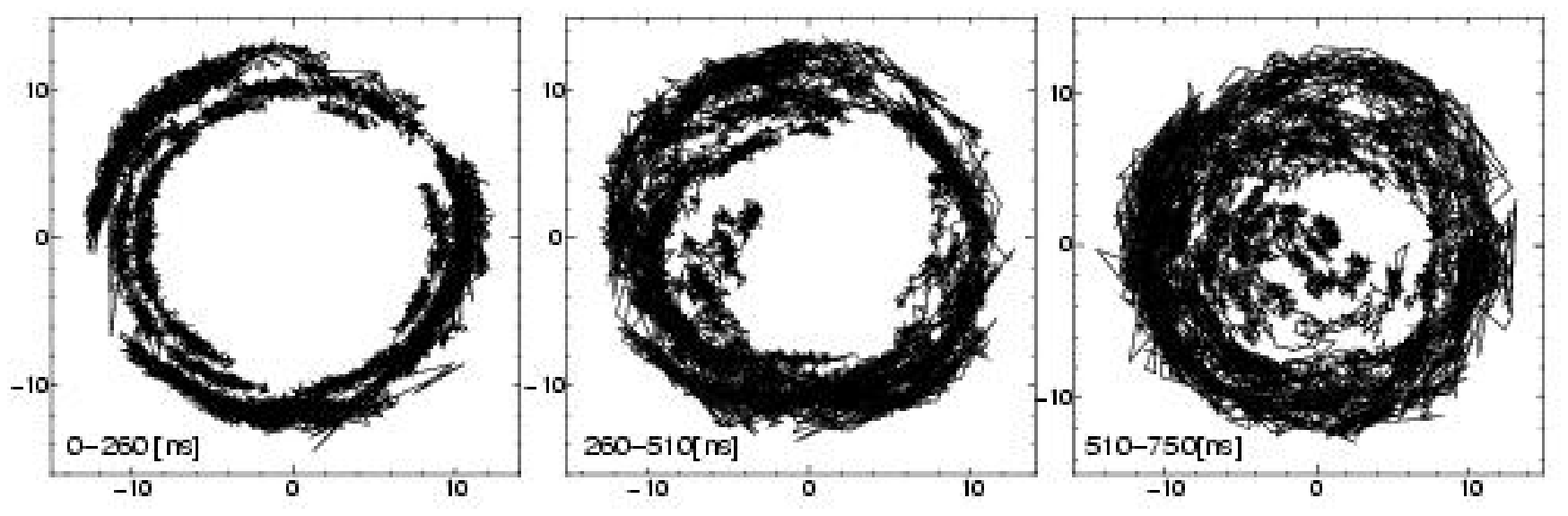}
\end{center}
{\bf Fig.11} 
Trajectories of solute atoms(B-atoms) during SA 
in a cluster $A_{47}B_{20}$($\alpha=1.1$). 
Initial configuration and initial temperature are 
the same as those in Fig.5. and Fig.4, respectively.
\end{figure}
During the initial stage of the process($\sim 260$[ns]) the 
B-atoms glide over the surface of a cluster.
In the next stage the atomic motion begins to contain
the component perpendicular to the surface 
and a diffusive motion into the cluster is activated($\sim 510$[ns]).
The trajectories of B-atoms spread 
over the whole cluster in the final stage.
In short, atomic gliding motion along the surface plays a 
role of a trigger to push solute atoms, which was initially 
located on the surface, into the inside of a cluster.
Successive snapshots in Fig.11 demonstrate that the diffusion along the 
surface induces the diffusion to the inside of cluster.\par
In addition, the diffusion into the solid cluster, 
which is responsible for the SA process, is quite different 
from the diffusion into the bulk solid in the following aspect.
A usual diffusion mechanism of atoms into the bulk is attributed to 
a hopping process which is mediated by {\it defects} or {\it vacancies} 
in the solid\cite{Shewmon}.
In contrast, a formation of defects in clusters is 
an extremely rare and its lifetime is very short, 
because defects are immediately pushed out to the surface.
Then the resulting hopping probability in the radial direction 
mediated by the defects is negligibly small.
Moreover, in our simulations, we could find no evidence 
suggesting that the vacancies inside the cluster play any role in the 
diffusion process in the radial direction of the cluster. 
Thus we conclude that the diffusion into the solid cluster, 
which is relevant for the SA process, is quite different from the 
diffusion process in the bulk solid. 
To elucidate the essential mechanism of the rapid diffusion
into the solid cluster will be discussed elsewhere in detail\cite{YIS}. 
In the present work, we only claim that, as discussed above, 
the frequent rearrangement of atoms on the surface of the cluster 
initiates the rapid diffusion in the direction perpendicular to the surface.
\section{Conclusion}
In the present paper we examined the presence and the absence 
of SA in a 2D binary microcluster in terms of an isoenergetic 
MD of a simple Morse model under different conditions of initial 
temperatures, sizes, and heat of solutions. 
One important advantage of our model is that 
the maginitude of negative heat of solution, 
which has been supposed to be he key parameter of SA in YM's 
experiment, can be well controlled by a single parameter.  
We confirmed that the heat of solution is similarly a 
driving force to form homogeneously mixed binary 
cluster in a very short time scale less than $1[\mu s]$   
in our model. 
Our main conclusions consist of the following three results.\\
\noindent (1) By changing the initial temperature of
the system, we found that SA occurs sufficiently below 
the melting temperature. 
The time required to complete SA becomes 
longer exponentially as the initial temperature $T_0$ decreases.
That is, it obeys an Arrhenius-like law,  
$\tau_{alloy} \propto \exp [\frac{T_{alloy}}{T_0}]$.\\
\noindent (2) We investigated the cluster size dependence of the
alloying time. As a result, we found observed that larger size 
cluster spend longer time to achieve SA. 
More precisely, the activation energy $T_{alloy}$ becomes 
larger with increase in the cluster size.
This result makes it clear that the quicker alloying 
surely occurs in smaller sized clusters.
By extrapolating the Arrhenius plot, we verified that
the alloying time is much less than $sec$ order at 
room temperature for a sufficiently small cluster. 
These numerical results qualitatively coincide with the experimental
observation by YM.\\
\noindent (3) By introducing quantites to probe 
fluctuating and rearranging 
properties of atoms composing a cluster, 
we found that the surface layer of a cluster is in a melting 
state even at the temperature much lower than the melting point 
of the cluster. 
The surface melting state is almost equivalent to a condition 
where atoms keep rearranging along the surface of the cluster.
Such an active surface motion is converted into the rapid diffusion
of solute atoms in the direction perpendicular to the surface
and results in a rapid SA. 
As far as such a diffusive motion assisted by surface melting is 
concerned, the rapid diffusion of the solute atoms 
into the solid {\it cluster} is quite different from the diffusion 
process in a {\it bulk} solid. 
It should be emphasized that the surface melting is the very 
origin to activate the radial diffusion process. 
As discussed in Sec.VI, 
the radial diffusion process is accerelated by a successive gliding 
motion of surface atoms, even if core part of a cluster is solid-like. 
The active motion of surface atoms is gradually converted into the
rapid radial diffusion by the frequent onset of a gliding motion of 
the surface atoms.  
A gliding motion as a collective atomic motion will be elaborated 
in detail elsewhere\cite{YIS}. \par
Before closing our conclusion, it is worth recalling 
three important factors, say substrate effect, dimensionality effect, 
and manybody effect, which are {\it not} taken into account in 
the present study.  
In fact we neglect the role of substrate 
which support a cluster and absorbs the heat 
accumulated in the alloying process.
By choosing isoenegetic MD simulation we assumed that the coupling 
strength between substrate and a cluster is very week and 
heat transfer from  cluster to substrate is considerably slow.
Although we roughly evaluate how fast is the heat transport in the 
appendix, there still exists a possibility that we underestimate 
the effect of a supporting substrate. 
If we emphasize a role of a substrate as a heat reservoir for clusters, 
an isothermal dynamics such as a Langevin simulation should be employed 
to trace the time evolution.
According to the Langevin dynamics, kinetic temperature of a cluster 
does not increase as SA proceeds, because the released heat of
solution is quickly absorbed by the substrate.
Thus, we are able to remove considerable temperature rise 
caused by negative heat of solution and explore the effect of 
temperature as purely as possible, distinguishing from the effect 
of negative heat of solution\cite{tkoba}.
As far as conclusions we present here concerned, 
the gross feature of SA in an isoenergetic condition is not much different 
from that in an isothermal one.\par
In the MD studies of bulk metals, manybody potential models 
are usualy employed to mimic interaction between metal atoms.
We examined an isoenergetic MD of SA with a manybody potential based 
upon the EAM, but we did not observe very significant differences 
from the present simulation\cite{SIS-EAM}.
From these facts we expect that, except for some detailed
apects, the peculiar features due to manybody potential 
do not essentially alter our results 
related to dynamics of SA process.
In this connection, it is worth noting that  
the presence of spontaneous mixing behavior has been also
reported for alkali halide microclusters(KBr-KCl system)\cite{Kaito}. 
The atomic interaction in alkali halide compound is 
well-described by Born-Meyer type pairwise potential, which is 
completely different from manybody interaction of metal atoms\cite{Tosi}.
For these reasons it is plausible to say that manybody effect,  
which is peculiar to metal, is not essential for the onset of SA. 
We employed a 2D model which is somehow special in a sense 
that it exhibits anomalous fluctuating properties near the 
melting point, which is similar to Kosterlitz-Thouless type 
transition\cite{Glaser}.
The direct outcome due to the confinement in 2D, not 3D,  
is also reported in comparison to the 3D model
with EAM\cite{SIS-EAM}.
Nevertheless, our preliminary results reveal that there 
are no significant differences between 
2D Morse model and 3D EAM model, as far as the materials
we examined here are concerned.\par
\begin{center}
{\bf Acknowledgment}
\end{center}
Authors thank H.Yasuda and H.Mori for their 
discussion and critical comments, and C.Satoko and T.Kobayashi 
for their helpful suggestions.  
One of authors (Y.S) thanks the financial 
support from JSPS Research for the Future Program in the Area 
of Atomic-Scale Surface and Interface Dynamics, and T.Yamabe 
for his continual encouragement. 
A part of the work is also supported by a Grant-in-Aid 
on Priority Areas, 'Chemistry of Small Manybody System',
from the Ministry of Education, Science, and Culture, Japan.  
\begin{appendix}
\section{HOW FAST DOES KINETIC ENERGY OF ATOMS TRANSFER FROM 
A CLUSTER TO THE SUBSTRATE?}
\indent In the present paper we assume that the heat given by 
formation of bonding between a deposited impurity atom
and cluster was released to the substrate slowly. 
In the appendix we evaluate how fast kinetic energy 
transfers from atoms in cluster to the substrate 
in terms of a simple one dimensional model.
As depicted in Fig.A.1 the cluster atoms are bounded to the substrate atom 
via harmonic potential, where mass of cluster atoms and 
substrate atoms are $M$ and $m$, respectively. 
\begin{figure}
\begin{center}
\epsfysize=3cm
\epsfbox{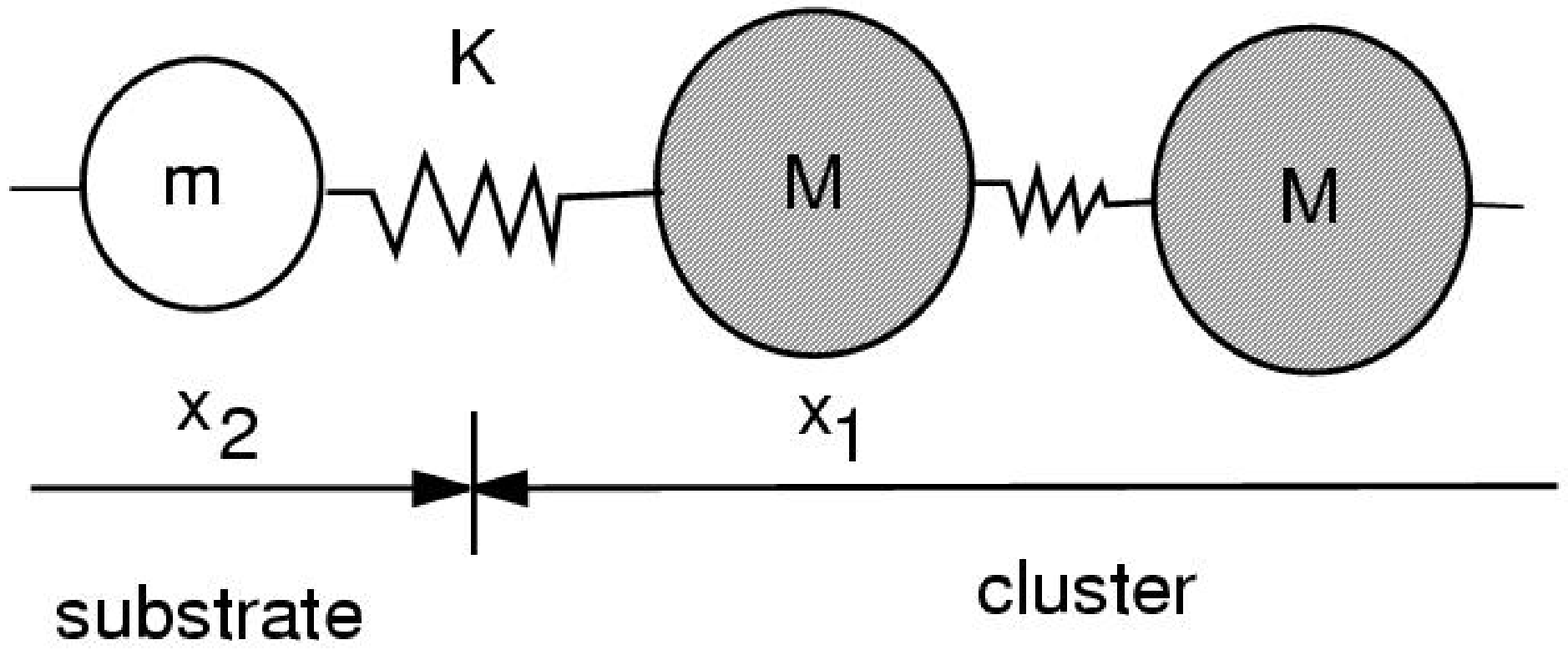}
\end{center}
{\bf Fig.A.1} A schematic picture of a 
substrate atom(white circle) interacting with a cluster atom 
(shaded circle) via harmonic potential.
\end{figure}
Spring constant of harmonic potential between a cluster 
and a substrate atom is denoted by $K$.
Atomic frequencies of a cluster atom and a substrate atom are 
$\omega_0$ and $\omega_1$, respectively. 
Langevin equation for two atoms locating interface between 
a substrate and a cluster is given by 
$$
\ddot{x}_1=-\omega_1^2 x_1+\frac{K}{M}(x_2-x_1)\\
\eqno{(A.1)}
$$
$$
\ddot{x}_2=-\omega_0^2 x_2-\gamma_0\dot{x}_2+f(t)+\frac{K}{m}(x_1-x_2)\\
\eqno{(A.2)}
$$
where $x_1$ and $x_2$ are positions of a substrate atoms and a cluster atom, 
$\gamma_0$ is a friction constant, $f(t)$ is an external random noise. 
$m$ and $M$ are mass of a substrate atom and a cluster atoms, 
$\omega_0$ and $\omega_1$ are vibrational frequencies of atoms inside 
the substrate and the cluster.
Laplace transform of these two equations is expressed as 
$$
z^2X_1-zX_1^{(0)}-\dot{X}_1^{(0)}=-\omega_1^2X_1+\frac{K}{M}(X_2-X_1)
\eqno{(A.3)}
$$
and 
$$
z^2X_2-zX_2^{(0)}-\dot{X}_2^{(0)}=
$$
$$
     -\omega_0^2X_2-\gamma_0zX_2+\gamma_0X_2^{(0)}+F(z)+\frac{K}{m}(X_1-X_2)
\eqno{(A.4)},
$$
where $X_1(z)\equiv {\cal L}[x_1(t)] =\int_0^{\infty}x_1(t)\exp(-zt) dt$, 
$X_2(z)\equiv {\cal L}[x_2(t)] = \int_0^{\infty}x_2(t)\exp(-zt) dt$, 
$F(z)={\cal L}[f(t)]$,
$X_1^{(0)}\equiv x_1(0)$ and $X_2^{(0)}\equiv x_2(0)$.
They lead us to the following expression for $X_2$;
$$
X_2=\frac{K}{m}\frac{X_1}{(z^2+\omega_0^2+\frac{K}{m}+\gamma_0 z)}+
\tilde{F}(z),
\eqno{(A.5)}
$$
where 
$$
\tilde{F}(z)=F(z)+(z+\gamma_0)X_2^{(0)}+\frac{\dot{X}_2^{(0)}}
{z^2+\omega_0^2+\frac{K}{m}+\gamma_0 z}.
\eqno{(A.6)}
$$
Our goal is to give a closed form to evaluate 
an effective friction constant for the variable $x_1$.
The substitution of (A.5) to (A.3) and its inverse Laplace transform 
yields,
$$
\ddot{x}_1(t)=-\omega_1^2 x_1-\frac{K}{M}x_1+
\frac{K}{M}\int^{t}_0 \theta (t-t')
x_1(t')dt' + g(t),
\eqno{(A.7)}
$$
where 
$$
\theta (t)\equiv \frac{\frac{K}{m}}{\tilde{\omega}}\exp(-\frac{\gamma_0}{2}t)
sin(\tilde{\omega}t)\\
\eqno{(A.8)}
$$
$$
\tilde{\omega}\equiv \sqrt{\omega_0^2+\frac{K}{m}-\frac{\gamma_0^2}{4}}
\eqno{(A.9)}
$$
$$
g(t)\equiv {\cal L}^{-1} [ \frac{K}{M} \tilde{F}(z)]
\eqno{(A.10)}
$$
By introducing a new variable $\Phi(t)=\int_t^{\infty} \theta(\tau) d\tau$,  
the third term of $(A.7)$ is rewritten as, 
$$
\int_0^t \theta(t-t')x_1(t')dt'=[\Phi (t-t')x_1(t')]_0^t-\int_0^t 
\Phi(t-t')\dot{x}_1(t')dt'. 
$$ 
Then, the resulting expression for $x_1$ is given by
$$
\ddot{x}_1=-\omega_1^2 x_1 -\frac{K}{M}(1-\Phi_0)x_1-\frac{K}{M}
\int_0^t \Phi(\tau-t')\dot{x}_1d\tau +\tilde{g}(t),
\eqno{(A.11)}
$$
where
$$
\tilde{g}(t)\equiv g(t)-\frac{K}{M}\Phi(t)x_1(0), 
\eqno{(A.12)}
$$
and $\Phi_0\equiv \Phi(t=0)$.
When comparing time scale of $\Phi(t)$ to that of  $\dot{x}_1$, 
it is easy to note that $\dot{x}_1$ oscillates with frequency $\omega_1$
and that temporal behavior of $\Phi$ is dominated by a frequency
$\tilde{\omega}$.
These values are determined by the frequencies of the substrate atoms 
and the cluster atoms, respectively. 
The value of $\omega_1$ is about 
$0.2\times 10^{14}[sec^{-1}]$ for 
Au, while  $\tilde{\omega}$ is roughly estimated as 
$3\times 10^{14}[sec^{-1}]$ for carbon graphite.  
We may say characteristic time scale for $\Phi$ and $\dot{x}_1$ 
is well separated. 
As a result, the third term of rhs of eq.(A.11) is simplified by extracting 
$\dot{x}_1$ out of integral.
\par
On the other hand, the explicit form for $\Phi$ is 
$$
\Phi(t)=\frac{K}{m \tilde{\omega}}\frac{e^{-\frac{\gamma_0}{2}t}}
{(\omega_0^2+\frac{K}{m})}(\tilde{\omega}\cos \tilde{\omega} t +
\frac{\gamma_0}{2} \sin \tilde{\omega} t),
\eqno{(A.13)}
$$
and the so-called frequency shift, say $\Phi_0$, is 
$$
\Phi_0=\frac{K}{m}\frac{1}{(\gamma_0^2+\frac{K}{m})}.
\eqno{(A.14)}
$$
In addition, if we assume to hold symmetric relation for $\Phi$, namely 
$\Phi(t)=\Phi(-t)$, and to extend upper limit of integral region 
respect to $\tau$ from $t$ to $\infty$, then $\int_0^{\infty} \Phi(t-\tau)
\dot{x}_1 d\tau=\dot{x}_1\int_0^{\infty} \Phi(\tau)d\tau$.
If we put $\beta=\int_0^{\infty} \Phi(\tau)d\tau$, then  
one can get the following expression from eqs.$(A.9)$ and $(A.13)$,
$$
\beta = \frac{K}{m}
\frac{\gamma_0}{(\omega_0^2+\frac{K}{m})^2}.
\eqno{(A.15)}
$$
By taking into account the relation $\gamma_0 \sim \omega_0$ and 
$\omega_0^2 \gg \frac{K}{m}$, it is possible to give 
the following relation,
$\beta \sim \frac{K}{m \omega_0^3}$.
Consequently, eq.(A.11) can be rewritten as 
$$
\ddot{x}_1=-\omega_1^2x_1-\frac{K}{M}x_1-\frac{K}{M}\beta\dot{x}_1
+\tilde{g}(t) 
\eqno{(A.16)}
$$
Then, we finally obtain a simple expression of the {\it effective} 
damping factor $\gamma$, 
$$
\gamma\equiv \frac{K}{M}\beta=\frac{K}{M}\frac{K}{m}\frac{1}{\omega_0^3}.
\eqno{(A.17)}
$$
It is safe to say that the damping factor $\gamma$ is small enough,  
since the ratio $\frac{\gamma}{\omega_1}$ is estimated 
as, 
$$
\frac{\gamma}{\omega_1}=( \frac{\sqrt{\frac{K}{M}}}{\omega_1})^2
(\frac{\sqrt{\frac{K}{m}}}{\omega_0})^2(\frac{\omega_1}{\omega_0}).
\eqno{(A.18)}
$$
Due to the frequency mismatch at the interface 
between the cluster atoms and the substrate atoms, one can easily 
show the following relations,
$
(\frac{\sqrt {\frac{K}{M}}}{\omega_1})^2 \sim \frac{1}{10},
(\frac{\sqrt {\frac{K}{m}}}{\omega_0})^2 \sim \frac{1}{10},
$
and 
$(\frac{\omega_1}{\omega_0}) \sim \frac{1}{10}$. 
Finally, we obtain $\frac{\gamma}{\omega_1}\sim 10^{-3}$, indicating that 
an energy transfer from a cluster atom to a substrate is sufficiently 
slow comparing to time scale of the atomic frequency of cluster atoms.
\end{appendix}

\end{multicols}

\begin{thebibliography}{99}
\bibitem{COM}
We examined the numerical simulation by EAM
with the set of parameters decided by Johnson for Au-Cu pair,
but we cannot recognize any particular difference from the Morse model  
as long as the time evolution process is concerned.
\bibitem{notice}
Concerning the value of $\alpha$, 
a density functional calculation for dimers, (Au-Au), (Au-Cu), 
and (Cu-Cu), yields
$\frac{\epsilon_{Au-Cu}}{\epsilon_{Au-Au}}=1.06$ 
and$\frac{\epsilon_{Au-A\ell}}{\epsilon_{Au-Au}}=1.37$ \cite{satoko}.
On the other hand the values of $\alpha$ is also evaluated by heat 
of solution 
and chohesive energy; $\alpha=1.05$ for $(Au-Cu)$ and $\alpha=1.2$ for
$(Au-A\ell)$.
\end{thebibliography}
\end{document}